\documentclass[3p]{elsarticle}
\usepackage{lineno,hyperref}
\usepackage{bm}
\usepackage{graphicx}
\usepackage{graphics}
\usepackage{amsmath}
\usepackage{cases}
\usepackage{amssymb}
\usepackage{booktabs}
\usepackage{multirow}
\usepackage{color}
\usepackage{xcolor}
\usepackage[normalem]{ulem}
\usepackage{tabularx}
\usepackage{siunitx}
\usepackage[caption=false]{subfig}
\usepackage{comment}
\usepackage{algorithm}
\usepackage{algpseudocode}
\usepackage{algorithmicx}

\journal{Computer Physics Communication}

\begin{document}

\title{GPU acceleration of an iterative scheme for gas-kinetic model equations with memory reduction techniques}

\author[strath]{Lianhua Zhu\corref{cor}}\ead{l.zhu@strath.ac.uk}
\author[strath]{Peng Wang}
\author[hust]{Songze Chen}
\author[hust]{Zhaoli Guo}
\author[strath]{Yonghao Zhang }\ead{yonghao.zhang@strath.ac.uk}

\address[strath]{James Weir Fluids Laboratory, Department of Mechanical and Aerospace Engineering, University of Strathclyde, Glasgow G1 1XJ, UK}
\address[hust]{State Key Laboratory of Coal Combustion, School of Energy and Power, Huazhong University of Science and Technology, Wuhan, 430074, China}
\cortext[cor]{Corresponding author}

\begin{abstract}
This paper presents a Graphics Processing Units (GPUs) acceleration method of an iterative scheme for gas-kinetic model equations. Unlike the previous GPU parallelization of explicit kinetic schemes, this work features a fast converging iterative scheme. The memory reduction techniques in this method enable full three-dimensional (3D) solution of kinetic model equations in contemporary GPUs usually with a limited memory capacity that otherwise would need terabytes of memory.  The GPU algorithm is validated against the DSMC simulation of the 3D lid-driven cavity flow and the supersonic rarefied gas flow past a cube with grids size up to 0.7 trillion points in the phase space. The performance of the GPU algorithm is assessed by comparing with the corresponding parallel CPU program using Message Passing Interface (MPI). The profiling on several models of GPUs shows that the algorithm has a medium to high level of utilization of the GPUs' computing and memory resources.  A $190\times$ speedup can be achieved on the Tesla K40 GPUs against a single core of Intel Xeon-E5-2680v3 CPU for the 3D lid-driven cavity flow.
\end{abstract}


\maketitle

\section{Introduction}
The dominant numerical method for rarefied gas flows remains to be the Direct Simulation Monte-Carlo (DSMC) method~\cite{birdMolecularGasDynamics1994, birdDSMCMethod2013}. However, with the significantly improved computing power nowadays, the deterministic methods including the discrete ordinate methods (DOM) and the discrete velocity method (DVM) which are proposed long before\cite{broadwellStudyRarefiedShear1964, yangRarefiedFlowComputations1995} to solve the Boltzmann equation or its model equations directly are rapidly gaining their popularity~\cite{aristovDirectMethodsSolving2001, mieussensSurveyDeterministicSolvers2014a}.
Compared with the DSMC method, these deterministic methods are free from statistic noise, thus are particularly preferable for low speed flows~\cite{aristovDirectMethodsSolving2001, soneMolecularGasDynamics2007,wuSolvingBoltzmannEquation2014,mengLatticeEllipsoidalStatistical2013,huangUnifiedGasKineticScheme2013,wuApparentPermeabilityPorous2017}. More importantly, the deterministic approach has greater flexibility in designing efficient numerical schemes, e.g., asymptotic-preserving schemes~\cite{jinAsymptoticPreservingAP2010, xuUnifiedGaskineticScheme2010, mieussensAsymptoticPreservingProperty2013, guoDiscreteUnifiedGas2013, guoDiscreteUnifiedGas2015, liuUnifiedGaskineticScheme2014a, xuDirectModelingComputational2015, zhuDiscreteUnifiedGas2016, zhuApplicationDiscreteUnified2017}, implicit schemes~\cite{mieussensDiscreteVelocityModel2000, titarevConstructionComparisonParallel2014, zhuImplicitUnifiedGaskinetic2016, zhuUnifiedGaskineticScheme2017, yangImplicitSchemeMemory2018, wangComparativeStudyDiscrete2018a} and high-order schemes~\cite{filbetHighOrderNumerical2003, alexeenkoHighOrderDiscontinuousGalerkin2008, wuThirdorderDiscreteUnified2018, suHighorderHybridizableDiscontinuous2018}.

However, even with the advanced numerical algorithms and simplified kinetic models, direct simulations of practical 3D problems by DVM are still computationally expensive regarding both the floating point operations and the memory requirement, due to the sheer number of grid points in the phase space (physical space and molecular velocity space). In most practical simulations especially involving high speed flows, massively parallel supercomputing facilities are indispensable~\cite{liGaskineticNumericalStudies2009, titarevConstructionComparisonParallel2014, liRarefiedGasFlow2015, dimarcoUltraEfficientKinetic2015,  liHighPerformanceParallel2016, titarevApplicationModelKinetic2017, dimarcoEfficientNumericalMethod2018, hoMultilevelParallelSolver2018}. Various techniques have been proposed to reduce the grid points, e.g., the adaptive local refinement for the spatial and velocity grids~\cite{kolobovUnifiedSolverRarefied2007, chenUnifiedGasKinetic2012, barangerLocallyRefinedDiscrete2014, brullLocalDiscreteVelocity2014, zabelokAdaptiveKineticfluidSolvers2015}. However, efficient parallelization on such non-regular grids can be a challenge especially for dynamic adaptive mesh refinement (AMR), where sophisticated load-balancing techniques are necessary.

The enormous computing cost of deterministic solvers for the gas-kinetic equations makes the adoption of emerging heterogeneous parallel computing platforms such as GPU and Intel Phi co-processors appealing. Programs executed on such platforms are written with specialized programming frameworks such as CUDA and OpenCL. Computational intensive parts of the algorithms are off-loaded to the accelerators which have onboard memory, and the computed results are copied back to the CPU memory. GPU, in particular, has achieved great success in the grid-based lattice Boltzmann method~\cite{fanGPUClusterHigh2004, tolkeTeraFLOPComputingDesktop2008, obrechtNewApproachLattice2011, mcclureNovelHeterogeneousAlgorithm2014, xuAcceleratedLatticeBoltzmann2017}  (a simplified DVM) and particle-based DSMC method~\cite{suLargescaleSimulationsMultiple2012, goldsworthyGPUCUDABased2014, jambunathanCHAOSOctreebasedPICDSMC2018}. For more general DVM simulations, GPUs have not been fully exploited\cite{frezzottiSolvingModelKinetic2011, frezzottiSolvingBoltzmannEquation2011, frezzottiDirectSolutionBoltzmann2011, klossSolvingBoltzmannEquation2010, aristovAccelerationDeterministicBoltzmann2011, zabelokGPUAcceleratedKinetic2012, rovenskayaNumericalInvestigationEffect2015}. Frezzotti et al.~reported a parallelization of a BGK-equation solver using an explicit DVM scheme and evaluated its performance on a Nvidia GTX-260 GPU that was released in 2009~\cite{frezzottiSolvingModelKinetic2011}. The reported speedup is over $300$ compared with a serial program on a contemporary CPU for the 2D lid-driven cavity flow with 3D velocity grid (2D3V configuration). However, the maximum grid size is limited to $160^2 \times 20^3$ as there is only $896$ MB memory on that GPU. Later, they extended the implementation to the full Boltzmann equation, achieving a speedup over 400~\cite{frezzottiSolvingBoltzmannEquation2011} on the same device with the grid size up to $192^2\times 18^3$. Kloss et al.~reported a GPU accelerated solver for the full Boltzmann equation with a conservative projection method and achieved a speedup up to 150 with a maximum grid size of $160^2\times 20^3$~\cite{klossSolvingBoltzmannEquation2010}.   Overstays and Crocke presented a similar GPU implementation that also solves the Boltzmann equation achieving an overall speedup around 50 on a low-end GPU~\cite{rovenskayaNumericalInvestigationEffect2015} with the maximum grid size being $128\times 64 \times 18 \times 18 \times 9$. Zabelok et al.~demonstrated a GPU acceleration of a 3D3V DVM module in their Unified Flow Solver (UFS) with the grid size up to $20000\times 32^3$ on a single GPU with the help of AMR in 3D physical space, and reported speedup of $20\sim 50$~\cite{zabelokGPUAcceleratedKinetic2012}.
The above literature reveals that except Zabelok's work~\cite{zabelokGPUAcceleratedKinetic2012},  all of the implementations are restricted to 2D problems, and the reason is the limited memory size on a single GPU.

Even though the global memory capacities on contemporary GPU devices have been growing, they are still relatively small ($4\sim 16$ GB) compared with the CPU nodes offering much fewer FLOPS. The memory size restriction can be technically mitigated with multi-GPU implementations, but the high-volume data transfer between multiple GPU devices may lead to new difficulties~\cite{zabelokMultiGPUKineticSolvers2014}. Another common feature of the above implementations is that they all use explicit discretization schemes for the convection term, which makes stencil computation well suited on GPUs.  However, for steady-state rarefied gas flows, we usually prefer an implicit or iterative DVM scheme which can reduce the computational time dramatically. For an implicit/iterative scheme, the data dependence in the physical space poses a new challenge for its parallelization, especially for the massively fine-grained parallel computing model like on GPU.

Here, we propose a GPU acceleration algorithm to solve a Boltzmann model equation~\cite{shakhovGeneralizationKrookKinetic1968} with memory reduction techniques, aiming to address the two issues discussed above, i.e., the GPU memory size barrier and the difficulty in parallelizing the iterative/implicit scheme on a GPU. In our algorithm, the memory requirement is reduced for both the molecular velocity and physical spaces. In the molecular velocity space, we employ the memory reduction technique proposed in Ref.~\cite{chenUnifiedImplicitScheme2017}. For a serial implementation of the scheme, we can only allocate the memory for a single discrete velocity and use it repeatedly for every other discrete velocity. For a parallelized implementation with decomposition in the molecular velocity space, each processor allocates storage for one discrete velocity, and the total memory occupation will be proportional to the number of parallel processors~\cite{chenUnifiedImplicitScheme2017, zhuThermallyInducedRarefied2017}. However, as the latest released GPUs contain several thousands of cores, the total memory occupation is still massive when parallelizing in the molecular velocity space, so we propose to further reduce the storage of physical space from 3D to 2D by employing the upwind character of this scheme. With these two techniques, the required memory space can be significantly reduced. In addition, the parallelization for the iterative scheme can be avoided by parallelizing in the molecular velocity space with each GPU core mapped to one discrete velocity, which is different from all the previous implementations~\cite{frezzottiSolvingModelKinetic2011, zabelokGPUAcceleratedKinetic2012}.

The remaining of the paper is organized as follows. In Section~\ref{sec:numerical_scheme}, we will solve one of the Boltzmann model equation, i.e., the Shakhov equation~\cite{shakhovGeneralizationKrookKinetic1968} to demonstrate our algorithm and explain how the memory occupation can be dramatically reduced in both the molecular velocity and physical spaces. In Section~\ref{sec:implementation} we give a simple introduction to the GPU programming paradigm using the CUDA framework and present the algorithms with the two main kernel functions. In Section~\ref{sec:validation}, the implemented GPU programs are verified with two 3D cases, i.e., the lid-driven cavity flow and the supersonic flow past a cube. In Section~\ref{sec:performance}, the parallel performance of the GPU programs is profiled on several different GPU models, and the speedup is compared with an MPI parallelization. The performance bottlenecks in various conditions are also identified. Finally, Section~\ref{sec:conlusions} concludes with our main findings and discussions for further development.

\section{Implicit kinetic scheme with the memory reduction technique} \label{sec:numerical_scheme}
\subsection{Shakhov equation}
The parallel implementation is based on an iterative scheme for steady state Boltzmann model equation. In this study, we use the Shakhov model~\cite{shakhovGeneralizationKrookKinetic1968} to demonstrate our implementation which is also applicable to other model equations. The governing equation of the velocity distribution function $f(\bm x, \bm \xi)$ reads as
\begin{equation}\label{eq:shak}
\bm \xi \cdot \bm \nabla f  = -\frac{f-f^S}{\tau},
\end{equation}
where $\tau$ is the relaxation time and is related to the local viscosity $\mu$ and pressure $p$ by $\tau = \mu/p$ to recovery the correct viscosity~\cite{sharipovRarefiedGasDynamics2015}.  $f^S$ is defined as
\begin{equation}\label{eq:feq}
f^S   = f^{M} \left[ 1 + (1-\text{Pr})\frac{\bm c \cdot \bm q}{5pRT}\left( \frac{c^2}{RT} -5 \right) \right ], \quad
\text{with} \quad f^{M} = \frac{\rho}{(2\pi RT)^{3/2}}\exp\left( - \frac{c^2}{2RT} \right),
\end{equation}
where Pr is the Prandtl number and $\bm c = \bm \xi - \bm U$ is the peculiar velocity with $\bm U$ being the hydrodynamic velocity.
For a monatomic gas, Pr equals to $2/3$.
The macroscopic variables such as the density $\rho$, velocity $\bm U$, temperature $T$, and heat flux $\bm q$ can be calculated from the moments of the distribution function,
\begin{equation}\label{mom}
\rho = \int f \text{d}\bm \xi, \quad  \rho \bm U = \int \bm \xi f \text{d}\bm \xi,\quad  \rho E = \frac{1}{2}\int \xi^2 \text{d} \bm \xi,\quad
\bm q = \frac{1}{2}\int \bm c c^2 f \text{d} \bm \xi,
\end{equation}
where $\rho E = 1/2 \rho U^2 + C_v T$ is the total energy with $C_v$ being the heat capacity [$(3/2)R$ for monatomic gases].
The pressure is related to the density and temperature by $ p = \rho RT$.

In DVM, the molecular velocity space is first discretized with a chosen 3D velocity grid  $\{\bm \xi_\alpha| \alpha={1,2,...,M}\}$.  The discrete form of the governing equation is then expressed as
\begin{equation}\label{eq:dbe}
\bm \xi_\alpha \cdot \bm \nabla f_\alpha
= - \frac{1}{\tau} [ f_\alpha - f_\alpha^S],
\end{equation}
where $f_\alpha$ is the distribution function of $\bm\xi_\alpha$ and $f_\alpha^S$ is the Shakhov-corrected equilibrium. The macro variables (moments) are evaluated by taking numerical integrations of $f_\alpha$ as follows,
\begin{equation}\label{eq:macros}
\rho = \sum_\alpha w_\alpha f_\alpha, ~ \rho \bm U = \sum_\alpha w_\alpha \bm \xi_{\alpha} f_\alpha, ~
\rho E  = \frac{1}{2}\sum_\alpha w_\alpha (\xi_{\alpha,x}^2 + \xi_{\alpha,y}^2 + \xi_{\alpha,z}^2) f_\alpha, ~
\bm{q} = \frac{1}{2}\sum_\alpha w_\alpha \bm c_{\alpha}
\left( c_{\alpha,x}^2 + c_{\alpha,y}^2 + c_{\alpha,z}^2 \right) f_\alpha,
\end{equation}
where $w_\alpha$ are the coefficients for the numerical quadratures.

\subsection{Iteration scheme with memory reduction technique}
Here we use a recently developed iteration scheme~\cite{chenUnifiedImplicitScheme2017} to discretize the discrete-velocity version of the governing equation, i.e., Eq.~\eqref{eq:dbe}. The advantage of this scheme is that we only need to store the distribution function value field of one discrete velocity, thus dramatically reducing the memory requirement from 6D to 3D for a full 3D problem. In the following, we first introduce the iterative scheme and then discuss its favorable parallelization approaches.

Equation~\eqref{eq:dbe} is solved using the following iterative scheme,
\begin{equation}\label{eq:iter_outer}
\bm \xi_\alpha \cdot \bm \nabla f_\alpha^{n+1}
= - \frac{1}{\tau^n} [ f_\alpha^{n+1} - f_\alpha^{S,n}],
\end{equation}
where $n$ denotes the index of iteration steps. The equilibrium part $f_\alpha^{S}$ calculated from macro variables is treated explicitly. Given $f_\alpha^{S,n}$ or equivalently the moments at the $n$th iteration step, and assuming the spatial gradient is evaluated with an upwind scheme, the $f_\alpha^{n+1}$ can be updated by a spatial sweeping sequentially along the characteristic direction for each discrete velocity~\cite{chenUnifiedImplicitScheme2017}. In the original paper, the authors proposed to rewrite the above formula in a delta form and use a central scheme to attain a second-order accuracy\cite{chenUnifiedImplicitScheme2017}, while introducing an inner sub-iteration loop in each iteration to recover the old distribution function. Introducing of the inner sub-iteration loop increases the overall computing time, and the more complex stencil computation makes the method more difficult to parallelize. In this work, we employ a first order upwind scheme for the gradient discretization. The accuracy of the first order scheme will be evaluated in Sec.~\ref{sec:validation}.

At the beginning of the iteration, $f_\alpha^{S,0}$ is set to the equilibrium state based on the initial (usually uniform) macro fields. Then the spatial field of the distribution function value of each discrete velocity is updated in a spatial sweeping manner on a Cartesian grid. The computational stencil is determined by signs of the discrete velocity components. For example, assuming we are currently process the discrete velocity $\bm \xi_\alpha$ with $\xi_{\alpha,\{x,y,z\}} > 0$,  when the spatial sweeping approaches to an inner cell with spatial index $(i, j, k)$, the distribution function $f^{n+1}_{\alpha, i, j, k}$ is updated from the following first order upwind discretization formula:
\begin{equation}\label{eq:sweeping_example}
\begin{split}
\xi_{\alpha,x} \left( \frac{f^{n+1}_{\alpha, i, j, k} - f^{n+1}_{\alpha, i-1, j,k} }{\Delta x_i}  \right)
+  \xi_{\alpha,y} \left( \frac{f^{n+1}_{\alpha, i, j, k} - f^{n+1}_{\alpha, i, j-1,k} }{\Delta y_j}  \right) \\
+  \xi_{\alpha,z} \left( \frac{f^{n+1}_{\alpha, i, j, k} - f^{n+1}_{\alpha, i, j,k-1} }{\Delta z_k}  \right)
= -\frac{1}{\tau^n_{i,j,k}}\left( f^{n+1}_{\alpha,i,j,k}  - f^{S,n}_{\alpha, i, j,k}\right).
\end{split}
\end{equation}
Because the same sweeping procedure for each discrete velocity can be executed individually without interaction with other discrete velocities, we can allocate the memory space for a single discrete velocity and process all the discrete velocities one by one using the same storage space. Once the whole filed of the distribution function of a specific discrete velocity is updated, its contributions are added up to the moments of the new iteration step. In this way, the same memory space is used repeatedly for all the discrete velocities.  After all the discrete velocities have been processed, the updating of new moments are also completed. The iteration can be repeated until it converges by monitoring the changes of moments. Boundary conditions are processed when the sweeping starts or approaches the domain boundaries.

By analyzing the data dependence of the algorithm, we can find the best parallelization strategy. Clearly, due to the directional spatial sweeping, each cell's stencil computation depends on its upwind cell information, so it is not straightforward to parallelize the sweeping in the physical space. In the traditional multi-core CPU approach, the classical domain decomposition in the physical space (multi-block parallelization) can partially solve this problem as the deterioration of convergence caused by the asynchronization between blocks is negligible due to the coarse granularity of the decomposition~\cite{titarevConstructionComparisonParallel2014, hoMultilevelParallelSolver2018}. But with the fine-grained GPU architecture, the asynchronization can dramatically deteriorate the convergence rate. One possible strategy is to decompose the domain among the diagonal sweeping wavefront but this approach requires careful consideration for load balance and the parallelism is limited by the relative small wavefront size~\cite{wassermanPerformanceScalabilityAnalysis2000, moustafaSharedMemoryParallelism2015, deakinImprovedParallelismScheme2016}. As the sweeping of each discrete velocity is independent of each other, the parallel computing in the molecular velocity space is much simpler. The communication only occurs at the stage for calculating moments. The parallelization in the molecular velocity space with Message Passing Interface (MPI) is straightforward in the same way as reported in Refs.~\cite{chenUnifiedImplicitScheme2017, zhuThermallyInducedRarefied2017}. The total memory will be increased with the number of parallel processes or threads. For MPI with distributed memory (multi-node) parallelization, the total memory requirement will not be an issue. However, for GPU parallelization with several thousands of tiny threads, the total memory occupation can be considerable and may exceed the memory capacity. We will show a memory reduction technique in the following section to tackle this problem.

\section{Parallel implementations using CUDA C}\label{sec:implementation}
\subsection{Brief introduction to GPU programming and overall considerations of the parallel implementation}
GPU enables massively parallel computing by supported by the advances in both hardware and software. A single GPU board nowadays contains thousands of light-weight streaming processors (SPs) or cores grouped into dozens of streaming multiprocessors (SMs) that can provide several teraflops computing power. It also comes with a hierarchical system of memories, registers, and caches to enable high-bandwidth and low-latency data feeding to the processors. Meanwhile, there are application programming interface (API) models that support general purpose computing on GPU devices by extensions of the currently popular programming language such as C++. The most widely adopted GPU programming API model is Nvidia's CUDA, the others include Khronos Group's OpenCL and Microsoft's DirectCompute). Since in this work we implement our algorithm using CUDA, here we briefly explain the related concepts in the CUDA environment.

CUDA employs a hierarchical thread organization structure in which the fine-grained computing tasks are mapped to a logical grid of threads which are grouped into thread blocks. Each thread blocks contains the same number of threads and are dynamically scheduled to the SMs. The threads execute specially defined functions called \emph{kernels} on the streaming processors. To realize the full potential of GPU computing power, the data/thread organization and kernel function implementation have to follow some principles, including:
\begin{itemize}
\item keeping memory access aligned in a half-thread wrap (consecutive 16 threads),
\item avoiding frequent data transfer between the CPU and GPU ends,
\item using the on-chip shared memory to make the threads in a thread block work together effectively,
\item limiting the usage of registers and shared memory in the kernels to ensure that enough thread blocks can reside on SMs (quantified as SM occupancy commonly).
\end{itemize}

The major floating-point operations in the iterative scheme described in Sec.~\ref{sec:numerical_scheme} come from the spatial sweeping and moment-evaluation procedures, i.e., Eqs.~\eqref{eq:sweeping_example} and~\eqref{eq:macros} respectively. For each procedure, we implement a kernel function, which is explained below.

\subsection{Spatial-sweeping kernel and further memory reduction in the physical space}
\begin{figure}[htbp]
\centering
\includegraphics[width=0.5\textwidth]{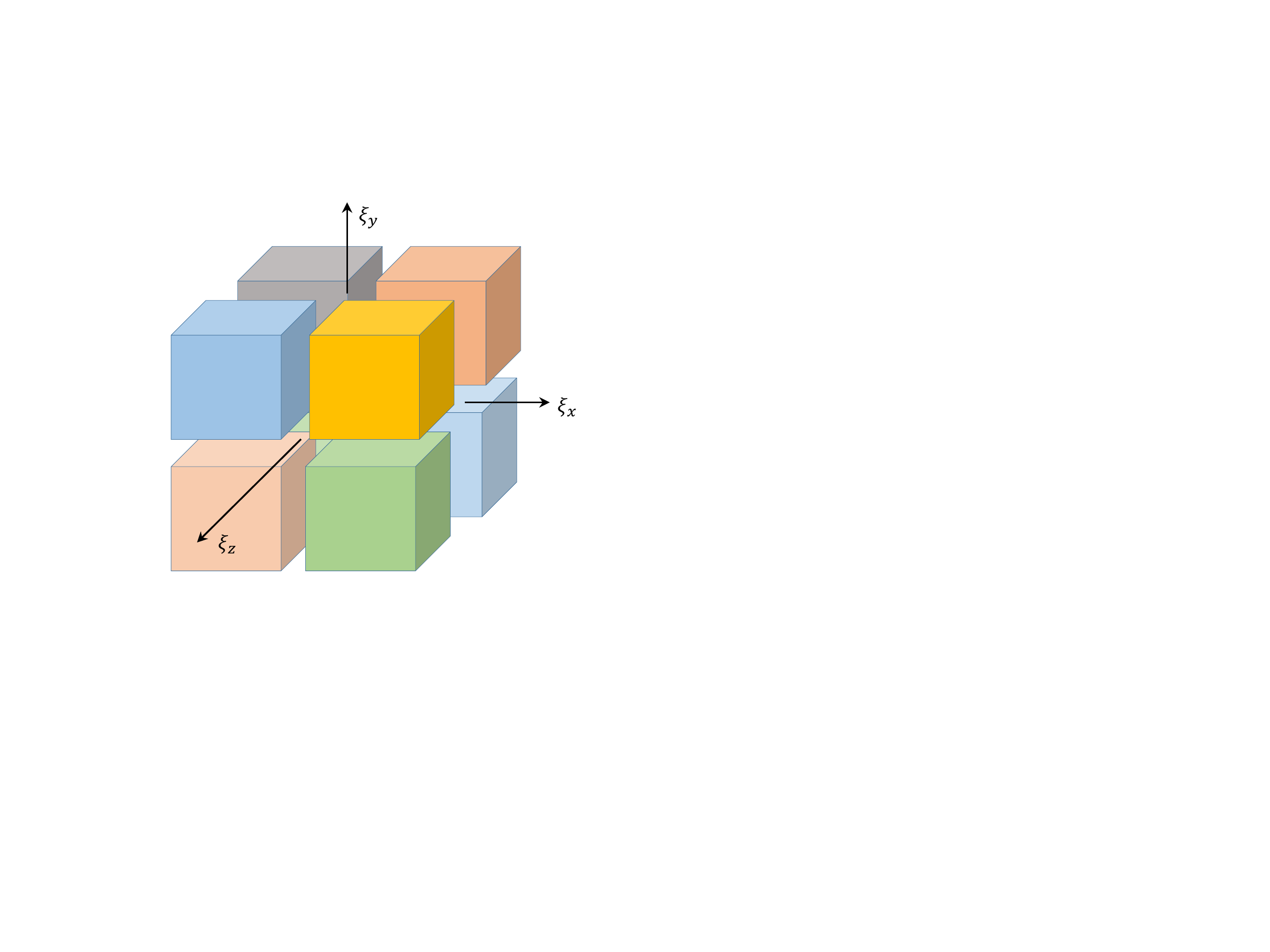}
\caption{The layout of the moment-evaluation kernel.}
\label{fig:dvg}
\end{figure}

As explained above, the spatial sweeping procedure is intrinsically sequential in the physical space while each discrete velocity is independent of other discrete velocities in the molecular velocity space.  So for the sweeping procedure, we map each thread to one discrete velocity. The thread grid size can be set to be a fraction of the total number of discrete velocity as we can process the discrete velocities with a batch-by-batch manner (see Fig.~\ref{fig:dvg} and the main iteration procedure in Algorithm~\ref{alg:main}). But it should still be large enough to make sure there are enough thread blocks to populate the SMs, which requires the grid size at least multiples of the number of SMs. One natural choice of the batch size is an octant of the discrete velocity grid as there are eight sweeping directions from the eight corners of the domain. The required memory is proportional to the thread grid size which may exceed the single GPU global memory capacity as modern GPU contains several thousands of cores. For example, for the Nvidia's Tesla K40 GPU which has 2880 cores and 12GB memory, when simulating a problem with the physical grid size of $128^3$ using single precision float, the modest estimation of the memory size exceeds $128^3 \times 2880 \times 4 /2^{30} = 22.5\text{GB} > 12\text{GB}$. So it is necessary to reduce the memory occupation further.

\begin{figure}[htbp]
\centering
\includegraphics[width=0.6\textwidth]{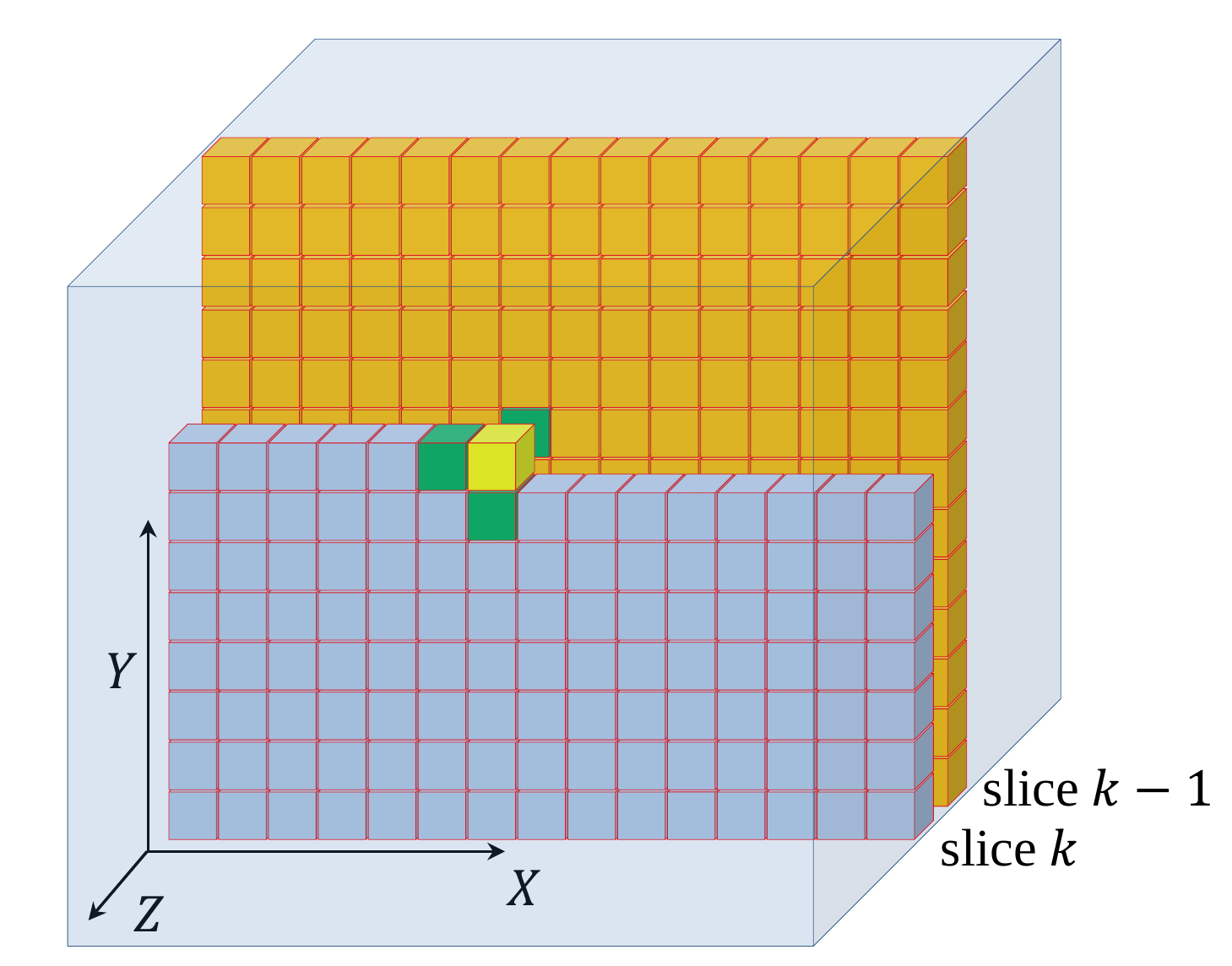}
\caption{Illustration of the spatial sweeping with the distribution functions of only two slices. The sweeping direction is $(0,0,0) \to (X^+, Y^+, Z^+)$. The highlighted cell in yellow color is the current cell being updated and the green cells are the upwind stencil cells (colorful online). The two slices are marching along the $0\to Z^+$ direction.}
\label{fig:sweep_illustration}
\end{figure}

\begin{figure}[htbp]
\centering
\includegraphics[width=0.8\textwidth]{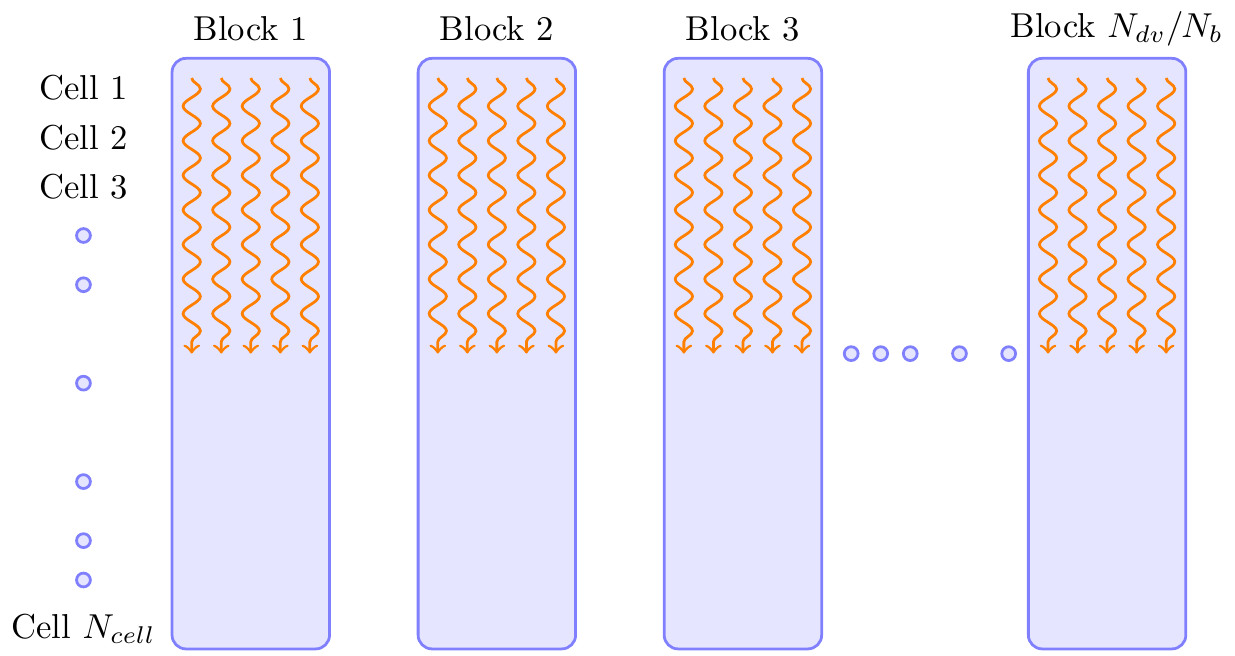}
\caption{The layout of the spatial-sweeping kernel, in which, $N_{cell}$ is the number of the cells in a 2D slice of the XY plane, i.e., $N_{cell} = N_x \times N_y$, $N_{dv}^{\theta}~( 0 < \theta < 8)$ is the number of discrete velocities in the $\theta$th batch, and $N_b$ is the size of the thread blocks.}
\label{fig:sweep_kernel}
\end{figure}

To this end, we propose to store the distribution function of only two-slices of cells by noticing the fact that when updating each cell's newer-step distribution function using a first order upwind scheme, all the stencil points are residing in the current and preceding slices. The storage space holds two successive slices of distribution function marching along the $Z$-direction as illustrated in Fig.~\ref{fig:sweep_illustration}. Such a configuration reduces memory consumption regarding the physical space dramatically from 3D to 2D. Together with the memory reduction technique described in Sec.~\ref{sec:numerical_scheme}, the current approach reduces the memory (single precision) from $4N^3M^3$ to $8N^2PQ$, where $N$ and $M$ stand for the physical and molecular velocity space grid numbers in one dimension respectively, and $P$ is the number of cores in a single GPU card, and $Q>1$ is a factor means the thread grid size is $Q$ times of the number of cores $P$. In this work, the thread grid size (batch size)  is set to be an eighth of the discrete velocity set, which is a natural choice considering that the sweeping directions in the eight-octant discrete velocities are different. But for the simulation with a huge discrete velocity set, the thread grid size can be even smaller to keep the overall memory within the GPU global memory limit. The layout of the thread grid is illustrated in Fig.~\ref{fig:sweep_kernel} and the kernel function implementation is detailed in Algorithm~\ref{alg:sweep_kernel}.

\begin{algorithm}
\caption{Main iteration procedure}
\label{alg:main}
\begin{algorithmic}[] 
	\Function{Iteration}{} 
	\For {$d \gets 0, 7$}  \Comment {Loop through 8 discrete velocity groups}
	\State Setting parameters for $d$th velocity group
	\If {$d > 3 $}  \Comment{Velocity groups above the $\xi_z=0$ plane in the molecular velocity space}
	\For {$k \gets 1, \texttt{NZ}$}\Comment{Sweeping along the $Z^+$ direction}
	\State \Call {sweep\_kernel}{$d$, $k$, \texttt{fSliceA}, \texttt{fSliceB}, \texttt{momentsOld}, \ldots}
	\State \Call {moment\_kernel}{$d$, $k$,  \texttt{fSliceA}, \texttt{fSliceB}, \texttt{momentsNew}, \ldots}
	\EndFor
	\Else   \Comment{Velocity groups below the $\xi_z=0$ plane in the molecular velocity space}
	\For {$k \gets \texttt{NZ}, 1$} \Comment{Sweeping along the $Z^-$ direction}
	\State \Call {sweep\_kernel}{$d$, $k$, \texttt{fSliceA}, \texttt{fSliceB}, \texttt{momentsOld}, \ldots}
	\State \Call {moment\_kernel}{$d$, $k$, \texttt{fSliceA}, \texttt{fSliceB}, \texttt{momentsNew}, \ldots}
	\EndFor
	\EndIf
	\EndFor
	\State \Call {swap\_moment\_kernel} {\texttt{momentsOld, momentsNew} } \Comment{Swap the moments variables}
	\EndFunction
\end{algorithmic}
\end{algorithm}

\algnewcommand\algorithmicswitch{\textbf{switch}}
\algnewcommand\algorithmiccase{\textbf{case}}
\algnewcommand\algorithmicassert{\texttt{assert}}
\algnewcommand\Assert[1]{\State \algorithmicassert(#1)}%
\algdef{SE}[SWITCH]{Switch}{EndSwitch}[1]{\algorithmicswitch\ #1\ \algorithmicdo}{\algorithmicend\ \algorithmicswitch}%
\algdef{SE}[CASE]{Case}{EndCase}[1]{\algorithmiccase\ #1}{\algorithmicend\ \algorithmiccase}%
\algtext*{EndSwitch}%
\algtext*{EndCase}%

\begin{algorithm}
\caption{Spatial-sweeping kernel function for the cavity flow.}
\label{alg:sweep_kernel}
\begin{algorithmic}
	\Function{sweep\_kernel}{$d$, $k$, \texttt{fSliceA}, \texttt{fSliceB},  \texttt{momentsOld}, \ldots}
	\State {$\texttt{tid} \gets \texttt{threadId.x}$} \Comment{Thread id.}
	\State {Define shared memory variables for moments \texttt{densL[NX2], vxL[NX2], vyL[NX2], vzL[NX2], \ldots } }
	\Switch{$d$} \Comment{Different entry points for 8 discrete velocity groups.}
	\Case{$0$} \Comment{For discrete velocity group 0, i.e., with $\xi_x <0,~\xi_y < 0, ~\xi_z < 0$}
	\If {k==\texttt{NZ}} \Comment{The slice near the boundary}
	\For {$j \gets 1, \texttt{NY}$}
	\For {$i \gets 1, \texttt{NX}$}
	\State {Update \texttt{fSliceA[j][i][tid]} as Maxwell boundary }
	\EndFor
	\EndFor
	\EndIf
	\For {$i \gets 1, \texttt{NX}$}  \Comment{The shaft near the boundary}
	\State {Update \texttt{fSliceB[NY+1][i][tid]} as Maxwell boundary}
	\EndFor
	\For {$j \gets 1, \texttt{NY}$}
	\If {$\texttt{tid} <\texttt{NX}+2$}
	\State{Collaboratively load moment variables from \texttt{momentsOld} to \texttt{densL[], vxL[], \ldots } .}
	\EndIf
	\State {Update \texttt{fSliceB[j][NX+1]} as Maxwell boundary} \Comment{The cell near the boundary}
	\For {$i \gets \texttt{NX},1$} \Comment{Stencil computation}
	\State {Calculate equilibrium distribution function \texttt{feq} using the moments in the shared memory.}
	\State {Calculate geometric, discrete velocity and $\tau$ related coefficients as \texttt{a, b, c, d, e}}
	\State {\texttt{fSliceB[j][i][tid] = (- a*fSliceB[j][i+1][tid]] - b*fSliceB[j+1][i][tid]}}
	\State{\texttt{\hspace{3.85cm}  - c*fSliceA[j][i][tid] + d[i]*feq ) * e}}
	\EndFor
	\EndFor
	\EndCase
	\Case{$1$}  \Comment{For discrete velocity group 1, i.e., with $\xi_x > 0,~\xi_y < 0, ~\xi_z < 0$}
	\State {Do similar loops as case 0}.
	\EndCase
	
	\State {\ldots}   \Comment{For discrete velocity groups $2 \to 6$.}
	
	\Case{$7$}  \Comment{For  discrete velocity group 7, i.e., with $\xi_x > 0,~\xi_y > 0, ~\xi_z > 0$}
	\State {Do similar loops as case 0}.
	\EndCase
	
	\EndSwitch
	\EndFunction
\end{algorithmic}
\end{algorithm}

\subsection{Moment-evaluation kernel}
\begin{figure}[htbp]
\centering
\includegraphics[width=0.8\textwidth]{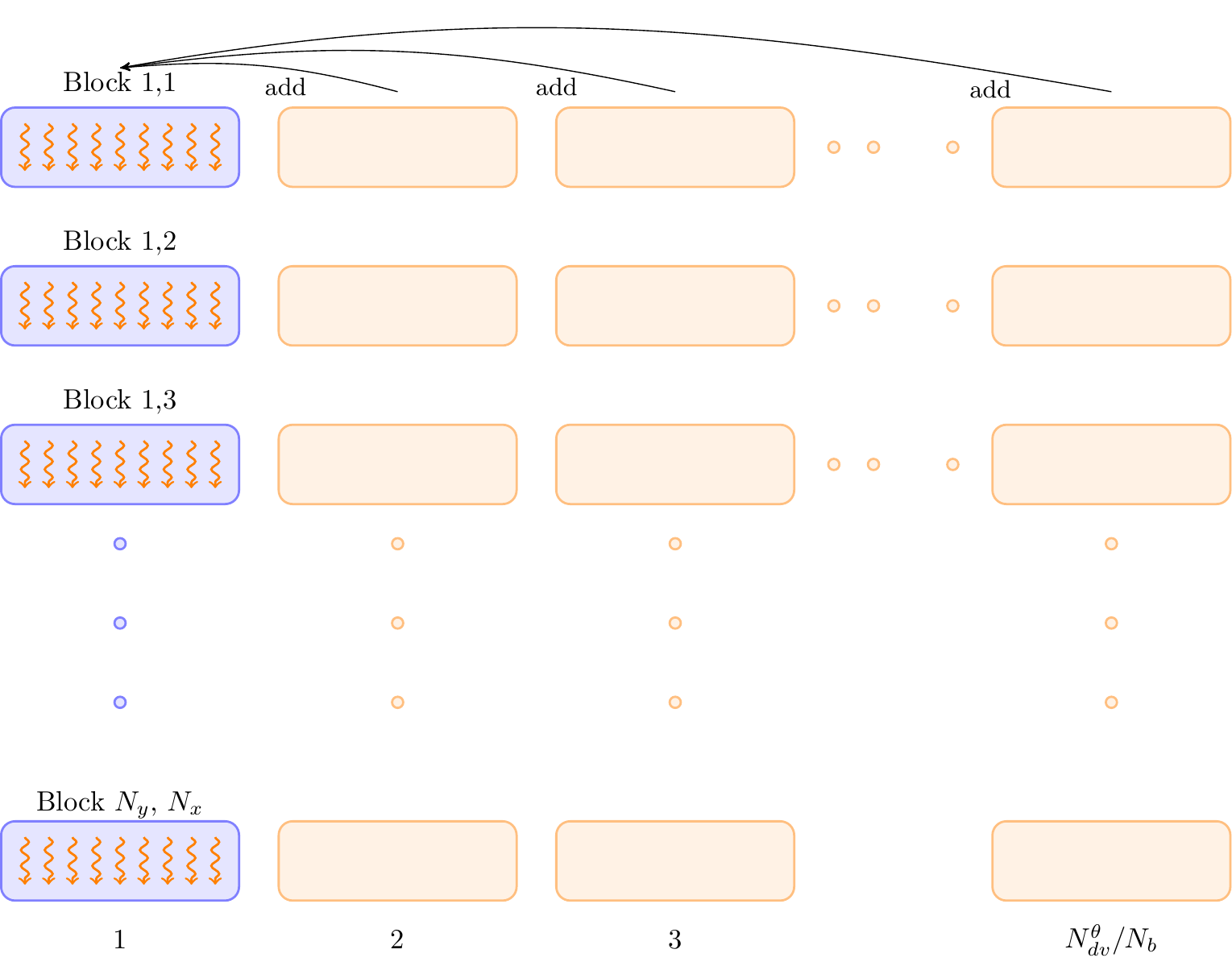}
\caption{The layout of the moment-evaluation kernel. Each thread block is mapped to one cell in the 2D slice. The threads in each block collaboratively add up the distribution function of all discrete velocities of the currently processed batch to the temporary moments. $N_{dv}^{\theta}~(0 < \theta < 8)$ is the number of discrete velocities in the $\theta$th batch. $N_b$ is the size of the thread blocks.}
\label{fig:moment_kernel}
\end{figure}

Each time after updates the distribution function of one slice of the cells, the moment-evaluation kernel adds up the contribution of these distribution functions to the temporary moment variables. As the summations are local operations in the physical space, we map each thread block to a single cell in the 2D slice, and the threads in one block work collaboratively using the shared memory to execute the binary reductions, i.e., sum up the distribution functions. The layout of the thread grids and blocks is shown in Fig.~\ref{fig:moment_kernel} and the outline of the moment-evaluation kernel is listed in Algorithm~\ref{alg:moment_kernel} in which the reduction operation using the shared memory is given in details.

\begin{algorithm}
\caption{Moment-evaluation kernel function for the cavity flow.}
\label{alg:moment_kernel}
\begin{algorithmic}
	\Function{moment\_kernel}{$d$, \texttt{k}, \texttt{fSliceA}, \texttt{fSliceB},  \texttt{momentsNew}, \texttt{\ldots}}
	\State {$\texttt{tid} \gets \texttt{threadId.x}$}
	\State {$\texttt{i} \gets \texttt{blockIdx.x}$} \Comment {Physical cell index in $X$ direction.}
	\State {$\texttt{j} \gets \texttt{blockIdx.y}$} \Comment {Physical cell index in $Y$ direction.}
	
	\State {Allocate shared memory for moments as \texttt{mom[8][BLK\_SIZE]} and initialize to zeros}
	\For {$b \gets 1, \texttt{dvGrpSize/BLK\_SIZE}$} \Comment{Add \texttt{BLK\_SIZE} discrete velocities per time.}
	\State {\texttt{dvId = b * BLK\_SIZE + tid}}
	\State {\texttt{fTmp = fSliceB[j][i][dvId]}}
	\State {\texttt{fSliceA[j][i][dvId] = fTmp}}
	\State {Accumulate \texttt{fTmp}'s contribution to \texttt{mom[0][tid]}, \ldots, \texttt{mom[7][tid]}}
	\EndFor
	\For {$s \gets \texttt{BLK\_SIZE/2}, 0$} \Comment {Parallel reduction with sequential addressing.}
	\If {$\texttt{tid} <  s$}
	\For {\texttt{m}$\gets 0, 7$}
	\State{$\texttt{mom[m][tid] += mom[m][tid+s] }$}
	\State{Synchronize threads}
	\EndFor
	\EndIf
	\EndFor
	
	\If {is thread 0} \Comment {Only one thread writes the data from the shared memory to the global memory.}
	\State{$\texttt{momentsNew} \gets \texttt{mom[][0]}$ }
	\EndIf
	
	\If{\texttt{i}, \texttt{j} or \texttt{k} near boundary}
	\State {Update solid boundary information}  \Comment{Extrapolate the wall densities needed for the next iteration step.}
	\EndIf
	\EndFunction
\end{algorithmic}
\end{algorithm}

\subsection{Boundary condition}
For the internal and external flows presented in the next section, we use the Maxwell diffuse boundary condition for the solid wall and the free-stream boundary condition for the external boundary.  Due to relatively simple geometries, in our implementation, the boundary condition treatments are embedded in the spatial-sweeping and moment-evaluation procedures.  In the Maxwell diffuse boundary, the velocity distribution function of the emitting (reflected) particles is:
\begin{equation}\label{eq:maxwell_wall}
f_{\alpha,w}^{n+1} = \frac{\rho_w^n}{(2\pi RT_w)^{3/2}}\exp\left[  -\frac{\xi_\alpha^2}{2RT_w} \right],
\end{equation}
where $\rho_w^n$ is the density determined by non-penetration condition:
\begin{equation}\label{eq:rho_wall}
\rho_w^n = -\sqrt\frac{2\pi}{RT_w}\sum_{\bm \xi_{\alpha} \cdot \bm n_w < 0}f_\alpha^{n} \bm \xi_\alpha \cdot \bm n_w.
\end{equation}
Note that $\rho_w^n$ is calculated using the last step ($n$) value of the distribution function of the particles coming to a wall because it's not possible to get other discrete velocities distribution functions at the newer step with the current iterative scheme. For an internal flow, the total mass in the domain will change slightly due to the mismatch of the iteration level in Eq.~\ref{eq:maxwell_wall}. Thus we scale the density of the whole field so that the total mass is unchanged. For the free streaming boundary condition, the distribution function of the particles entering into the computational domain is set to be the equilibrium distribution based on the free stream gas state. When the spatial sweeping starts from a boundary, the upwind cells' distribution function is determined by above-mentioned method. When it reaches to a wall, the discrete velocity's contribution to the density flux is accumulated in the moment-evaluation procedure.

\section{Validation of GPU implementation}\label{sec:validation}
In this section, we validate our GPU algorithm implementations on two 3D cases, i.e., the lid-driven flow in a cubic cavity and the supersonic gas flow past a cube. We first make sure the GPU programs give the identical results as the CPU versions, then we only compare the results of our GPU programs' results with the DSMC solutions obtained by the open source dsmcFoam solver\cite{scanlonOpenSourceParallel2010}. The main purpose is to assess the relative accuracy of the first order DVM scheme compared with the DSMC method which can also be viewed as a first order accurate method. In all of the simulations, the gas media is Argon, and its viscosity changes with the temperature as $\mu = \mu_\text{ref}(T/T_\text{ref})^{\omega}$ with $\omega=0.81$. The reference viscosity $\mu_\text{ref}$ at the reference temperature $T_\text{ref}$ is calculated as
\begin{equation}\label{eq:mu_ref}
\mu_\text{ref} =\frac{15\sqrt{\pi}p_\text{ref}\lambda_\text{ref}}{(5-2\omega)(7-2\omega)\sqrt{2RT_\text{ref}}},
\end{equation}
where $p$ and $\lambda_\text{ref}$ are the reference pressure and mean-free-path. All the solid walls are assumed to be fully diffuse boundaries described by Eq.~\eqref{eq:maxwell_wall} and~\eqref{eq:rho_wall}.

\subsection{Lid-driven rarefied flow in  a cubic cavity}
\begin{figure}
\centering
\includegraphics[width=0.4\textwidth]{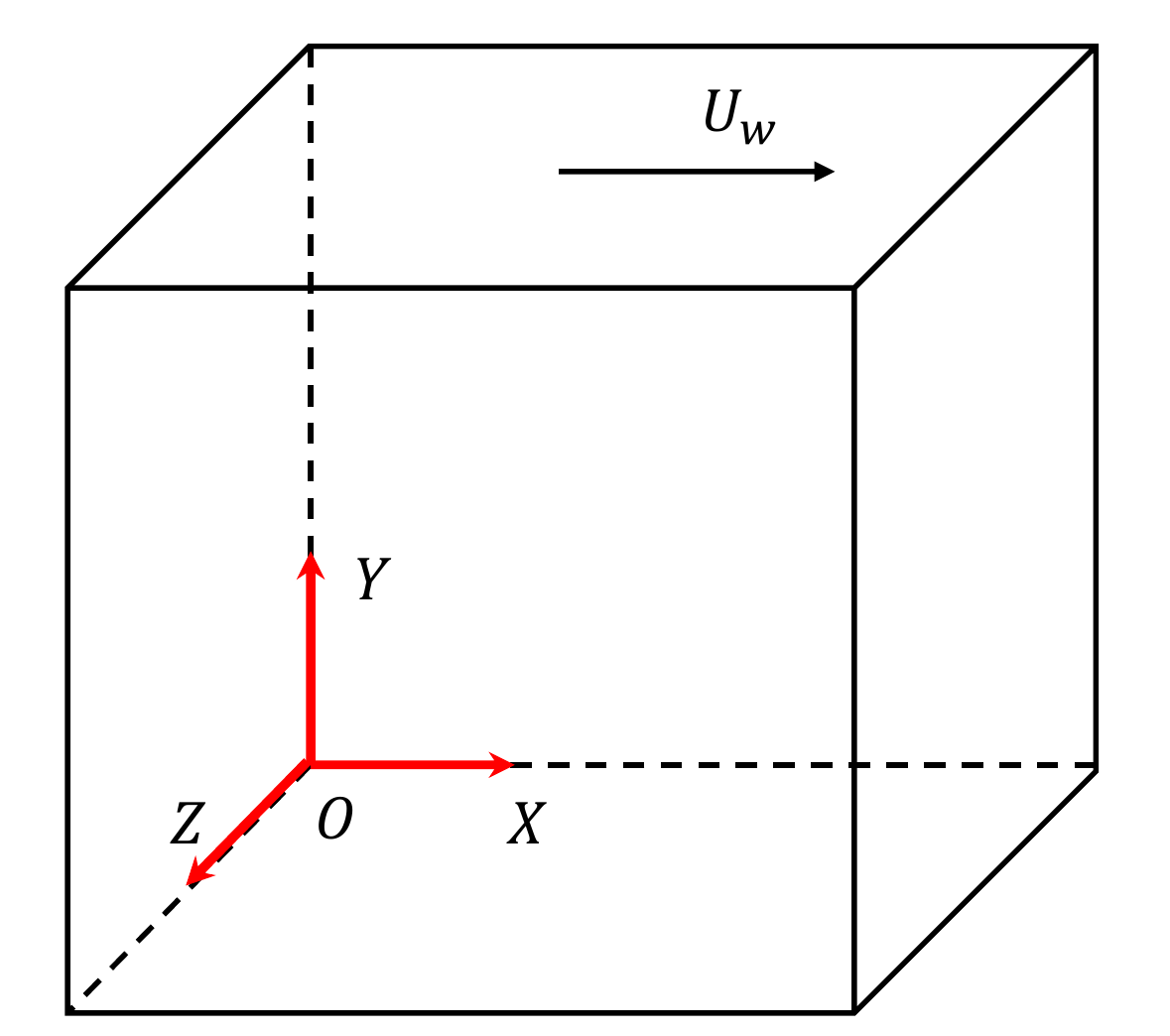}
\caption{Illustration of the three dimensional lid-driven cavity flow. The lid (top wall) moves in the $X+$ direction with a constant velocity $U_w$. All of the walls are maintained at a uniform temperature of 1.}
\label{fig:ldc}
\end{figure}

The first case is the lid-driven rarefied gas flow in a cubic cavity as illustrated in Fig.~\ref{fig:ldc}. The inner walls are maintained at a uniform temperature of $T_w=273.15K$. The lid (upper wall) of the cavity moves in the $X^+$ direction with a constant velocity of $U_w=0.1\sqrt{2RT_w}$. Based on the initial constant pressure and the side length of the cavity, the Knudsen is $1.0$. For this case, we use the same uniform Cartesian grids with $64^3$ cells in the physical space for both DVM and DSMC simulations. In the DVM, the velocity grid has $64^3$ points uniformly distributed in the range of $[-4\sqrt{2RT_w}, 4\sqrt{2RT_w}]^3$, and the trapezoidal rule is used for the evaluation of moments. The convergence criteria of the iteration in DVM is that the $L_2$ normal of the relative changes of the density, momentum components, and energy fields between two successive iteration steps are less than $10^{-9}$. The DVM simulation takes 41 minutes with 36 iteration steps on the Tesla K40 GPU. In the DSMC simulation, there are 20 particles in each cell on average. The DSMC simulation is run without sampling in the initial 20,000 steps and then run with sampling for further 600,000 steps to output the results with takes 141 hours with 128 CPU cores.

\begin{figure}[htbp]
\centering
\includegraphics[width=0.5\textwidth]{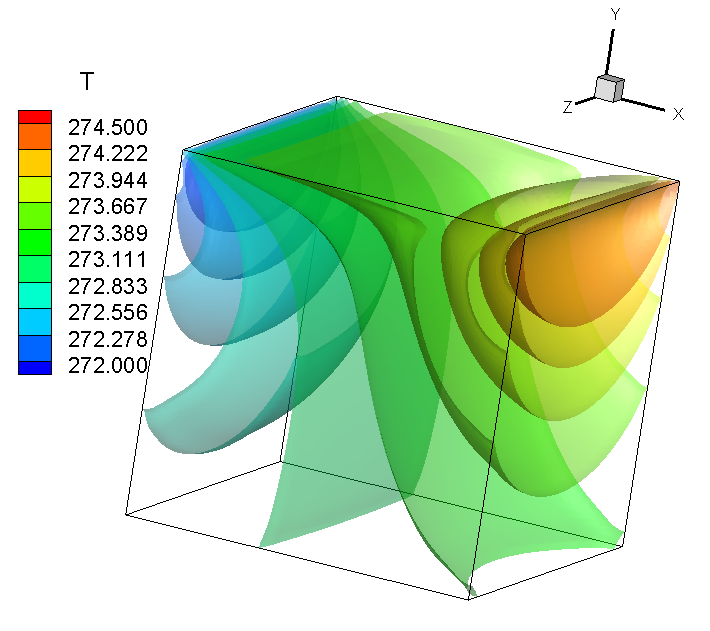}~
\includegraphics[width=0.5\textwidth]{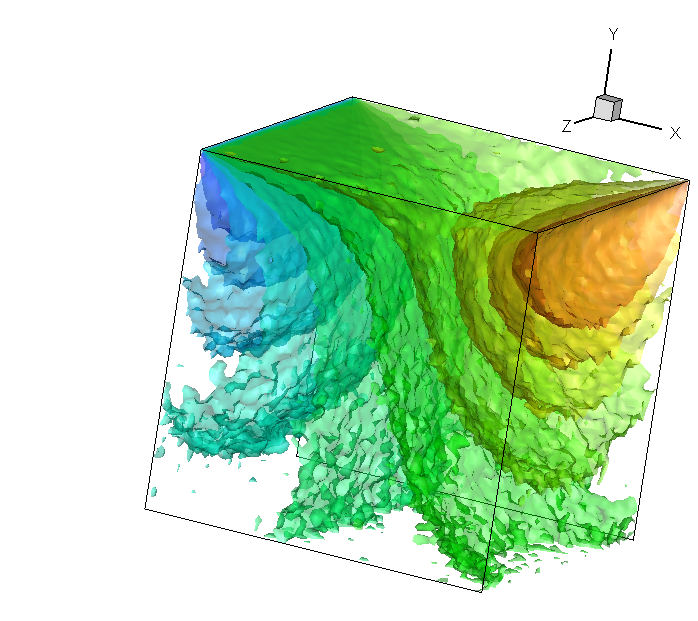}
\caption{Comparison of the temperature iso-surfaces predicted by the GPU accelerated DVM and the DSMC method. Spatial grid in both the DVM and DSMC are $64^3$, the velocity grid in the DVM simulation is $64^3$, and Kn is 1.}
\label{fig:cavity_iso}
\end{figure}

\begin{figure}[htbp]
\centering
\subfloat[]{\includegraphics[width=0.6\textwidth]{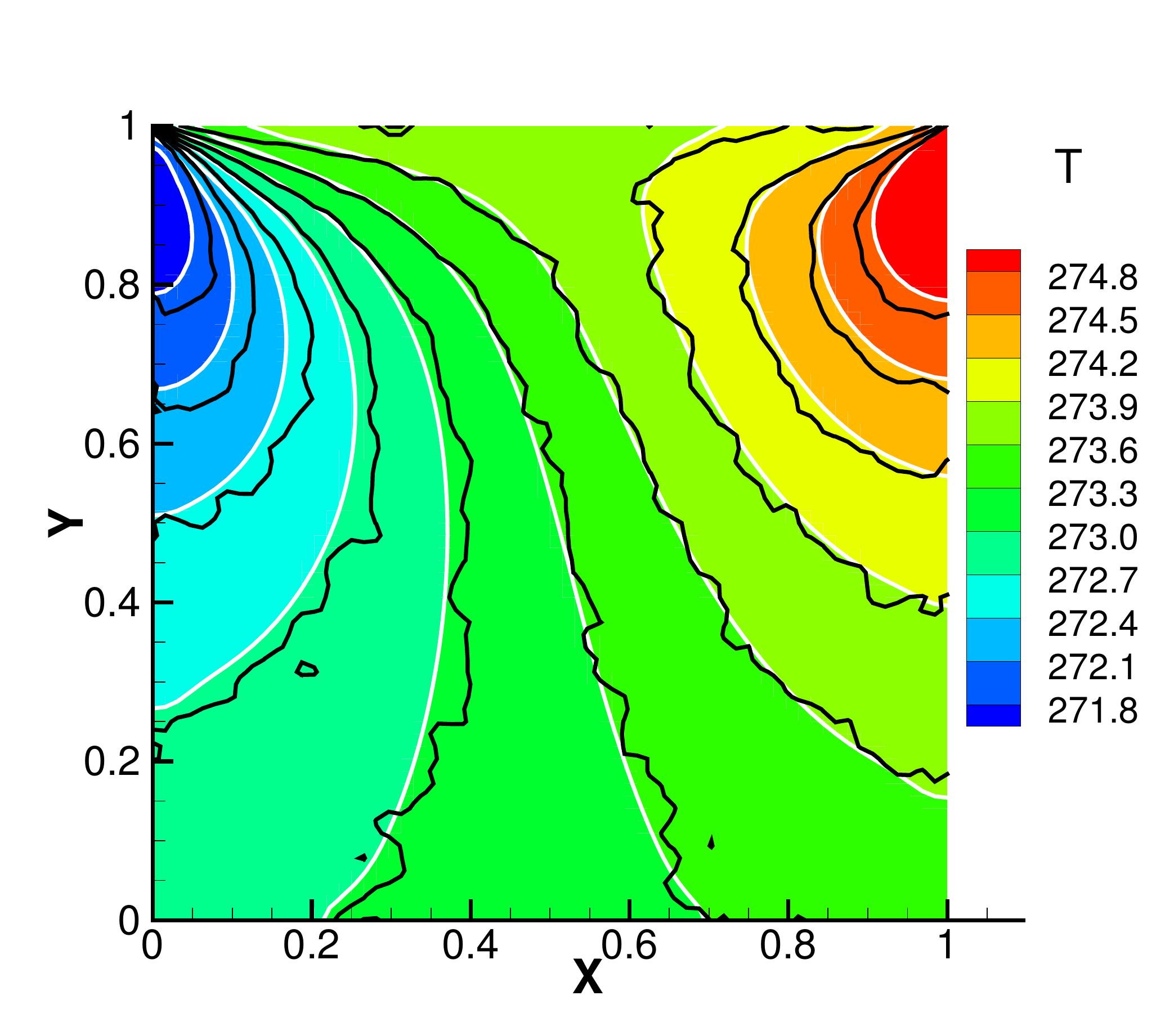}}~
\caption{Temperature distribution on the Z-symmetric plane of the cavity flow predicted by the GPU accelerated DVM and the DSMC method.}
\label{fig:cavity_contour}
\end{figure}

Figure~\ref{fig:cavity_iso} compares the temperature iso-surfaces from the DVM and DSMC solutions. We can observe overall good agreement between the two sets of results even though there are still significant noises in the DSMC solution. The temperature field in the $X$-$Y$ symmetric plane is shown in Fig.~\ref{fig:cavity_contour}(a) which confirms the overall agreement between the DVM and DSMC solutions regarding the temperature field. Figure~\ref{fig:cavity_contour}(b) shows the $X$/$Y$- component velocity profiles along the vertical/horizontal central line of the cavity. The velocity profile obtained with a finer mesh ($128^3$) has also been included for comparison. The minor difference between the profiles with two grid resolutions suggests that the first order DVM with a physical grid size of $64^3$ can already give a mesh independent velocity field at $\text{Kn} = 1$.

\begin{figure}
\centering
\includegraphics[width=0.44\textwidth]{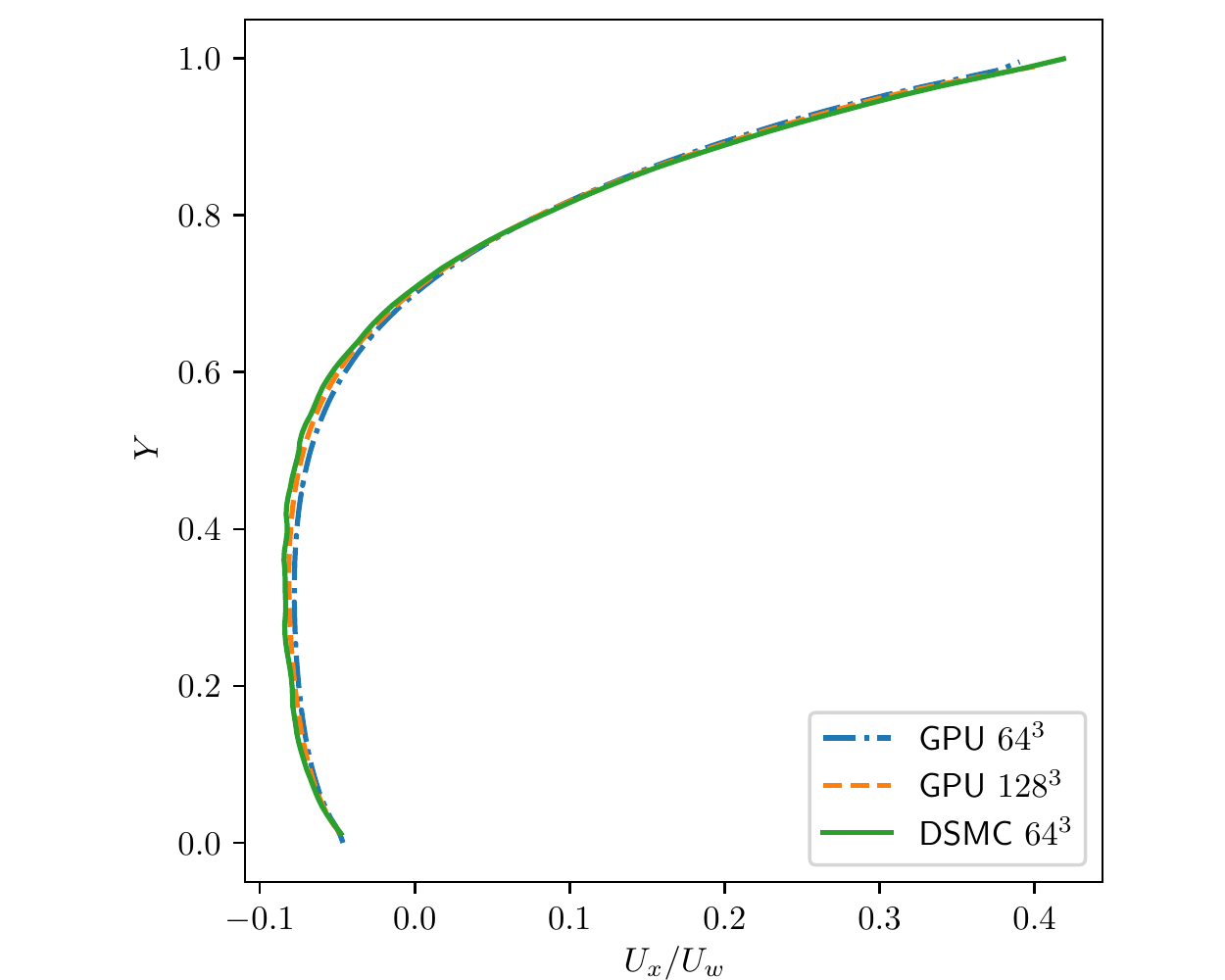}~
\includegraphics[width=0.44\textwidth]{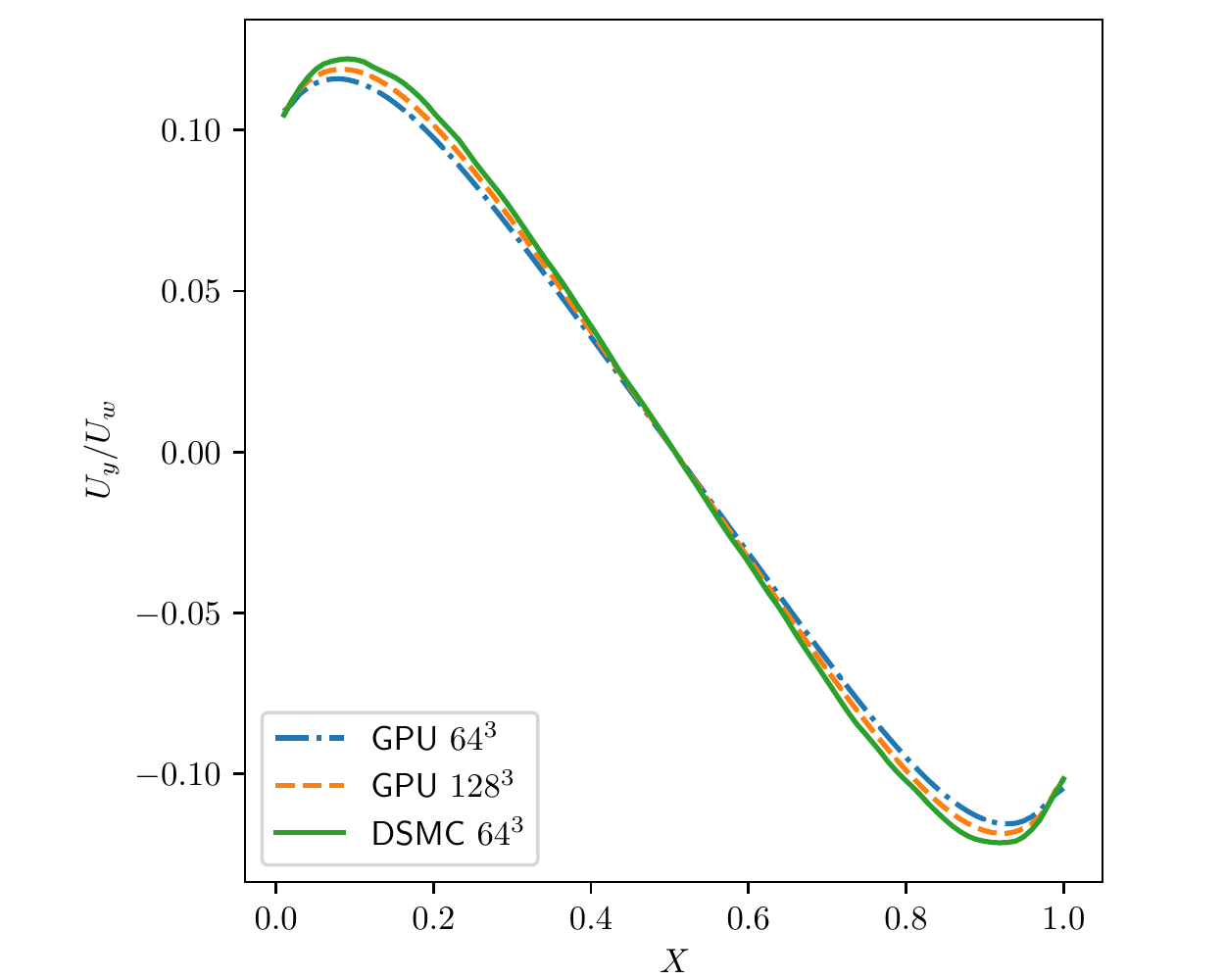}~
\caption{Horizontal velocity ($U_x$) profiles along the vertical central line (a) and vertical velocity ($U_y$) profiles along the horizontal central line (b) predicted by the DSMC and the GPU accelerated DVM simulation with different physical grid size.}\label{fig:aa}
\end{figure}

\subsection{Rarefied gas flow past a cube}
\begin{figure}[htbp]
\centering
\includegraphics[width=0.8\textwidth]{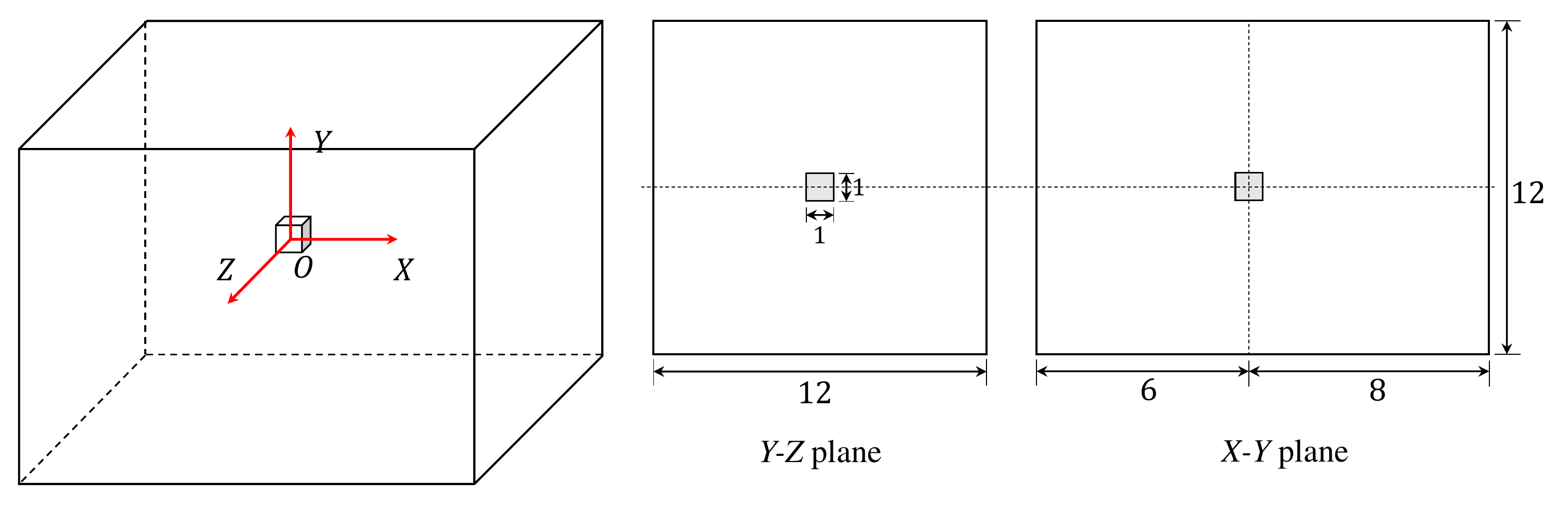}
\caption{Schematic diagram of the rarefied gas flow past a cube.}
\label{fig:cube_illustration}
\end{figure}

The second case is a supersonic rarefied gas flow past a cube with $\text{Ma}=2$ and $\text{Kn} = 1$. The cube's size is $1^3$ while the computational domain's size is $14\times 12\times 12$. The center of the cube is located at $(0,0,0)$ while the computation domain is given by Fig.~\ref{fig:cube_illustration}. The surface of the cube is maintained at 273K. In the DVM simulation, the whole domain is solved for convenience, while in the DSMC simulation, we simulate only a quadrant of the domain due to the two-fold symmetry. The full mesh size is $181\times 181 \times 191$. The cell size expands with a cell-by-cell ratio of 1.02  in the front and lateral sides of the cube and 1.03 at the rear of the cube. The velocity grid size is uniform with $48^3$ points in the range of $[-4\sqrt{2RT_w}, 4\sqrt{2RT_w}]^3$  and a trapezoidal rule is used to calculate the moments. The DVM simulation takes approximately 20 hours with 41 iteration steps on the Tesla K40 GPU. In the DSMC simulation, each cell has 50 particles on average and the time step size is 2.0e-7s. The sampling begins from 1,000 steps and continues for 68,000 steps which takes 128.5 hours on 128 CPU  cores.

\begin{figure}[htbp]
\centering
\subfloat[]{\includegraphics[width=0.5\textwidth]{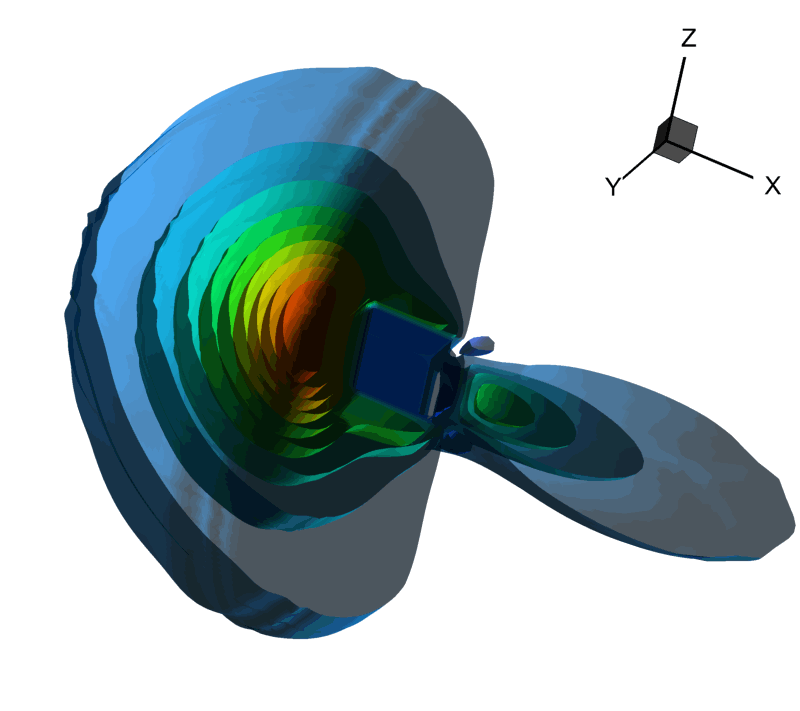}}
\subfloat[]{\includegraphics[width=0.5\textwidth]{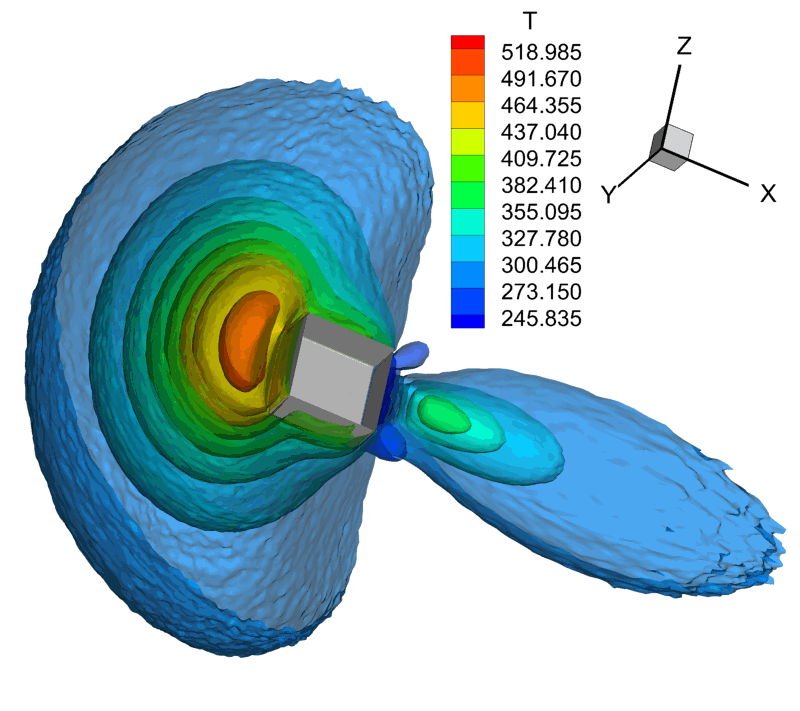}}
\caption{Temperature iso-surfaces predicted by the GPU accelerated DVM simulation and the DSMC simulation.}
\label{fig:cube_T_iso}
\end{figure}

\begin{figure}[htb]
\centering
\includegraphics[width=0.45\textwidth]{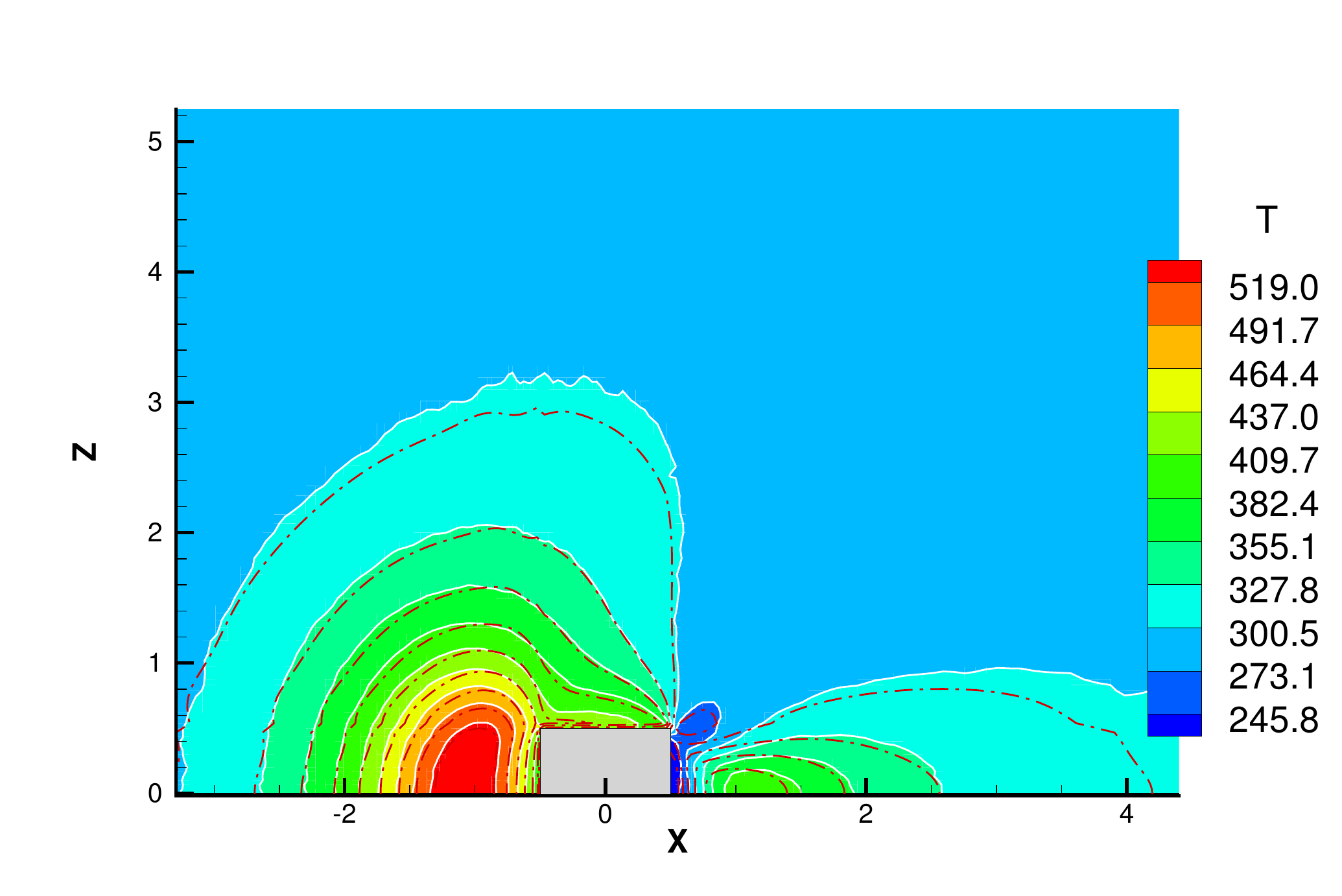}
\includegraphics[width=0.45\textwidth]{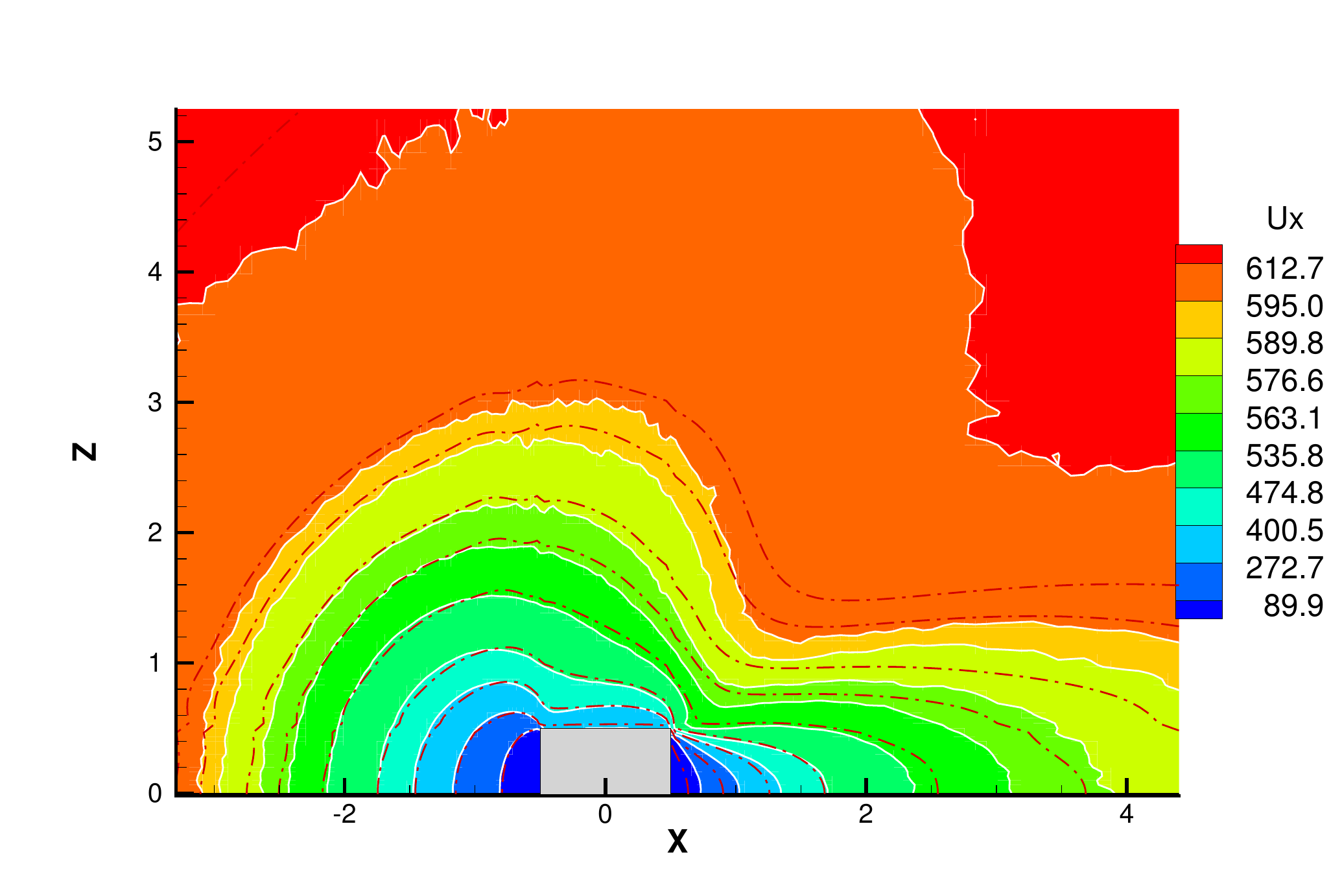}\\
\includegraphics[width=0.45\textwidth]{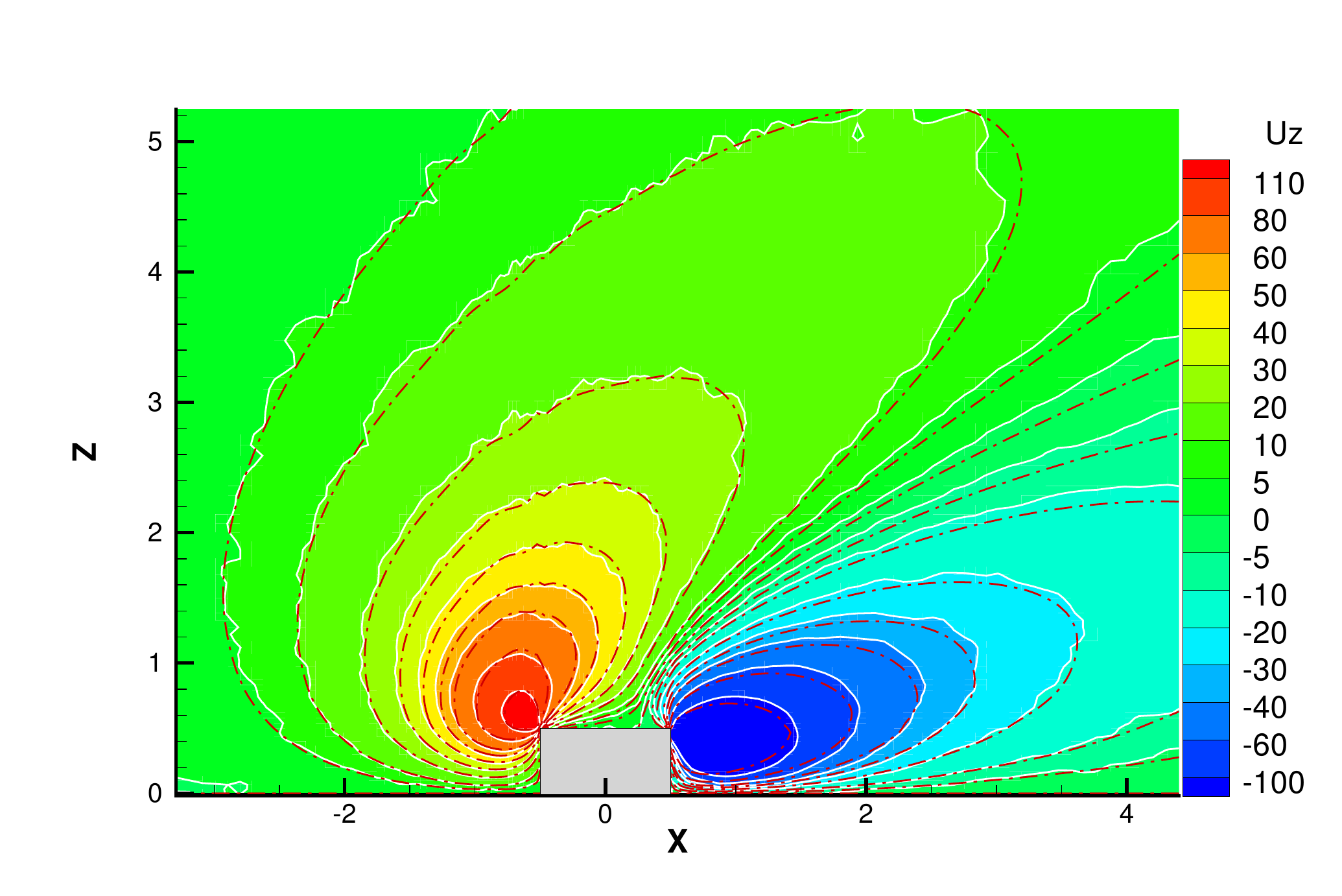}
\includegraphics[width=0.45\textwidth]{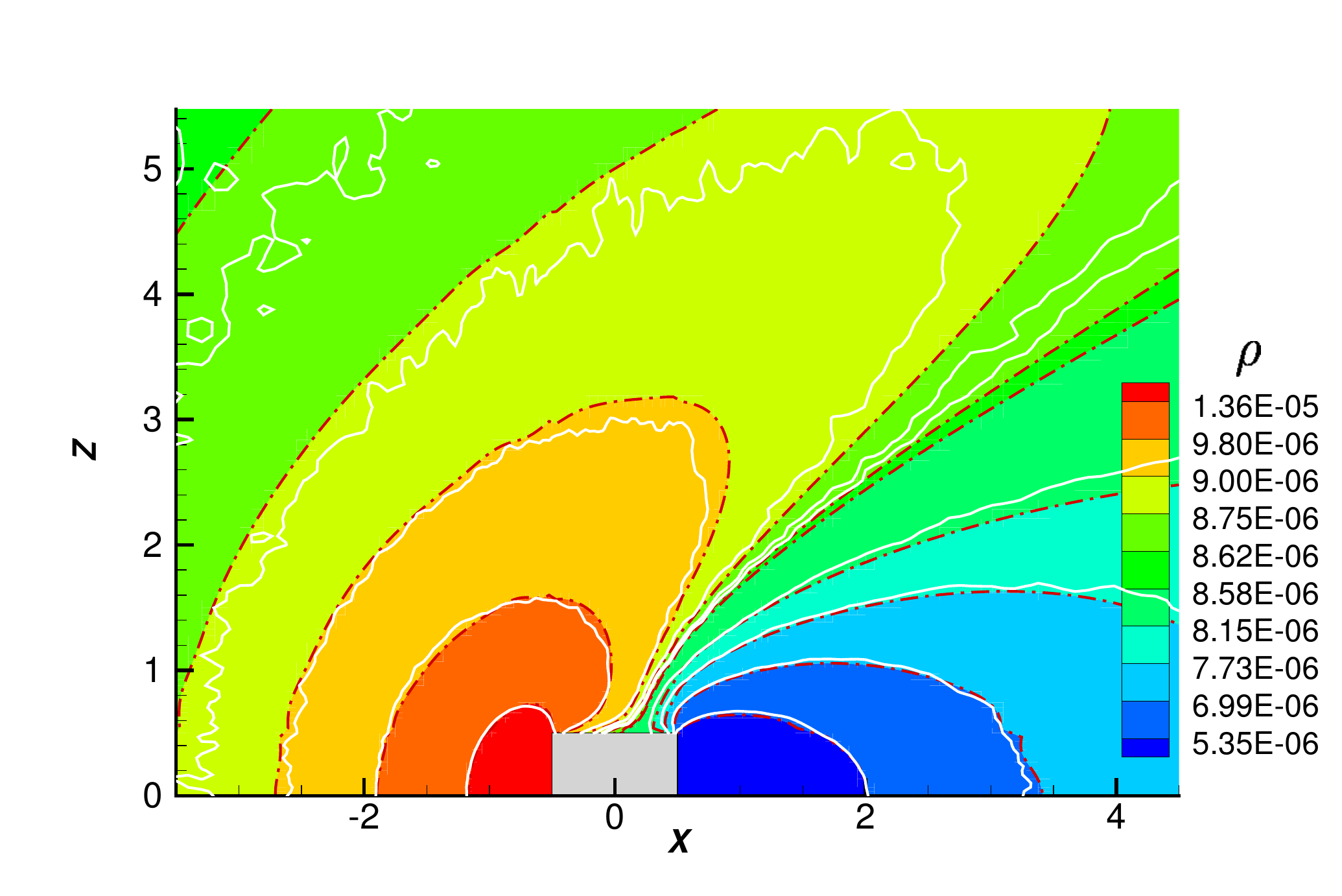}
\caption{Comparison of the temperature (a),  X-component (b),  Z-component velocity  (c) and density (d) distributions in the symmetry plane.  The dashed red lines denote the DVM result. The colored background and the solid white lines represent the DSMC solution.}
\label{fig:cube_slice_contours}
\end{figure}

The temperature iso-surfaces are presented in Fig.~\ref{fig:cube_T_iso}, from which an overall agreement between the DVM results and the DSMC solutions can be found. Figure~\ref{fig:cube_slice_contours} shows the detailed comparisons of the temperature, density and velocities distributions on the symmetry plane of the computation domain. The overall agreement indicates that the DVM prediction near the surface of the cube is satisfactory.

\section{Performance profiling and comparison with MPI parallelization strategy}\label{sec:performance}
In this section, we analyze the parallel computing efficiency and scalability of our GPU algorithm.  We also compare the speedups with an MPI parallel implementation of the same iterative scheme on a cluster of CPU nodes. All of the testings are based on the 3D lid-driven cavity flow with single precision computations. The GPU cards used in the platform include a Tesla K40 GPU, a Tesla K80 GPU and a Quadro M2000 GPU. The major specifications of the three GPU models are listed in Table~\ref{tab:gpu_specs}. Quadro M2000 GPU is based on a newer architecture called Maxwell while K40 and K80 are based on the older Kepler architecture. Quadro M2000 GPU has less streaming cores, but the multiprocessors can operate at a much higher frequency than K40/K80 GPU so it can also offer over a teraflops for single precision operations. The commercial Tesla K80 GPU card actually contains two GK201 GPUs, while in this study we only use a single GPU on it and all the data listed in Table~\ref{tab:gpu_specs} are the values of a single GK201 GPU. Our  program is developed with CUDA C++ and is compiled  using the Nvidia CUDA Toolkit (version 9.2) without the aggressive optimization flags such as \texttt{-prec-div=false, -prec-sqrt=false} or \texttt{-use\_fast\_math}.  The host compiler is Intel's C++ compiler (version 15.0).

In order to investigate the relative computing speed of the GPU program, we developed a CPU program parallelized with MPI and written in C++. The MPI-CPU program is compiled with the same compiler and Intel MPI library with compiling flag \texttt{-xHost} which will enable advanced arithmetical instructions such as AVX2 and FMA. The program runs on an in-house cluster equipped with a 56Gbps InfiniBand network and each node comprises two Xeon E5-2680v3 (Haswell) @2.5GHz CPU. The MPI-CPU program employs the domain decomposition in the molecular velocity space, which is straightforward to implement based on a serial program as the only data communication occurs at the moment-evaluation stage which can be easily implemented using the \texttt{MPI\_Allreduce} subroutine.

\subsection{Parallel speedups and comparison with the MPI CPU program}
Firstly, we investigate the overall speedup of the GPU algorithm on different GPU devices. The average computing time for a single iteration step of the cavity flow case with the various grid size of physical and molecular velocity space is measured on different platforms. The results are presented in Table~\ref{tab:iteration_time}, and the corresponding speedups against the MPI 1-core  are shown in Fig.~\ref{fig:speed_up}. The measured GPU global memory consumptions are also listed in Table~\ref{tab:iteration_time} to demonstrate the advantage of our memory reduction techniques. For the two largest grids, i.e., $64^3\times 128^3$ and $64^3 \times 128^3$, the global memory occupations are over 4 GB and therefore unable to run on the M2000 GPU. It should be remarked that if we don't use the memory reduction techniques above, for the two cases the most conservative estimation (storing only one float variable on each grid point in the phase space) of the memory occupation can be as high as $128^3\times 64^3 \times 4/10^{12} = 2.2$ TB.

Several interesting patterns can be observed from the table and chart. Firstly, when the velocity grid is $64^3$ or larger, both K40 and K80 achieve speedups around $190$.  It is much higher than the cases with the $32^3$ velocity grid which is only around $100$. This contrast can be easily explained. As there are over two thousand cores on K80/K40 GPUs, when simulating the cases with the $32^3$ velocity grid, the thread grid size of the \texttt{sweepSlice} kernel is only $16^3=4096$, meaning that there is simply not enough parallelism for the GPUs to fulfill their computing potential. Secondly, comparing speedups among three GPUs, we can found that K40 is about $20\%$ faster than K80 GPU, this is reasonable as the number of cores and memory bandwidth of K40 are relatively higher than those of K40, while their other configurations are almost the same.  M2000 GPU performs better than the other two for the cases with the $32^3$ velocity grid even though it has much less streaming processors and smaller global memory bandwidth. This means that for these cases, the computing power and high memory bandwidth on K40/K80 are not well used. Lastly, we consider the MPI CPU program performance. When running with 96 CPU cores, the MPI parallelization can achieve a speedup about $70$, which means the strong scaling parallel efficiency is around $73\%$, which is typical for the collective communications of the \texttt{MPI\_Allreduce} when reducing a large chunk of data.

\begin{table}[htbp]
\centering
\caption{Specifications of the three different GPUs used in the performance evaluation. For all of the GPUs, the GPU auto-boost feature is turned off and the fixed SM frequencies are sustainable in all the workloads. The measured global memory accessing bandwidth is obtained from the \texttt{bandwidthTest} program in the NVIDIA\_CUDA\_SAMPLES.
	\small\textsuperscript{*} The parameters for K80 GPU card are for a one of its two GK201 GPUs inside.}
\begin{tabular}{lrrrrrrr}
	\toprule
	GPU Model & \multicolumn{1}{l}{Number } & Device & \multicolumn{1}{l}{SM } & Bandwidth  & Peak single  & Max shm & Max reg \\
	& \multicolumn{1}{l}{of SP} & memory & \multicolumn{1}{l}{ frequency} & theoretical & precision & per SM & per SM\\
	\midrule
	Quadro M2000 & 768   & 4 GB   & 1088 MHz   & 106 GB/s & 1.786 TFLOPS & 96 KB & 64 K \\
	Tesla K40 & 2880  & 12 GB  & 745 MHz   & 288 GB/s & 4.29 TFLOPS & 48 KB & 64 K  \\
	Tesla K80* & 2496  & 12 GB  & 745 MHz   & 240 GB/s & 4.11 TFLOPS &  112 KB & 128 K   \\
	\bottomrule
\end{tabular}%
\label{tab:gpu_specs}%
\end{table}%

{\renewcommand{\arraystretch}{1.2}%
\begin{table}[htbp]
	\centering
	\caption{Average computing time (s) for a single iteration step and GPU global memory occupations on different platforms for the lid-driven cavity flow case. The last two cases on the M2000 GPU are not available due to its insufficient global memory capacity.}
	\begin{tabular}{rrrrrrrrrrrrrrr}
		\toprule
		\multicolumn{1}{c}{Physical} & \multicolumn{1}{c}{Velocity} &       & \multicolumn{3}{c}{CPU MPI Program} &       & \multicolumn{5}{c}{GPU Program}       &       & \multicolumn{1}{c}{GPU } \\
		\cmidrule{4-6}\cmidrule{8-12}    \multicolumn{1}{c}{grid $M$} & \multicolumn{1}{l}{grid $N$} &       & \multicolumn{1}{r}{1 core} &       & \multicolumn{1}{r}{96 cores} &       & \multicolumn{1}{r}{M2000} &       & \multicolumn{1}{r}{K40} &       & \multicolumn{1}{r}{K80} &       & \multicolumn{1}{c}{memory} \\
		\cmidrule{1-2}\cmidrule{4-4}\cmidrule{6-6}\cmidrule{8-8}\cmidrule{10-10}\cmidrule{12-12}\cmidrule{14-14}
		$32^3$ & $32^3$ &       & $53.50$ s &       & $0.68$ s &       & $0.48$ s &       & $0.55$ s &       & $0.57$ s &       & 75 MB \\
		$32^3$ & $64^3$ &       & $424.56$ s &       & $4.92$ s &       & $3.29$ s &       & $2.15$ s &       & $2.52$ s &       & 347 MB \\
		$64^3$ & $32^3$ &       & $429.32$ s &       & $6.20$ s &       & $3.77$ s &       & $4.33$ s &       & $4.42$ s &       & 206 MB \\
		$64^3$ & $64^3$ &       & $3097.39$ s &       & $43.29$ s &       & $26.33$ s &       & $16.57$ s &       & $19.80$ s &       & 1205 MB \\
		$64^3$ & $128^3$ &       & $24366.40$ s &       & $358.41$ s &       & ---   &       & $127.56$ s &       & $132.84$ s &       & 9242 MB \\
		$128^3$ & $64^3$ &       & $25020.61$ s &       & $371.02$ s &       & ---   &       & $131.04$ s &       & $156.47$ s &       & 4703 MB \\
		\bottomrule
	\end{tabular}%
	\label{tab:iteration_time}%
\end{table}%

\begin{figure}
	\centering
	\includegraphics[width=0.85\textwidth]{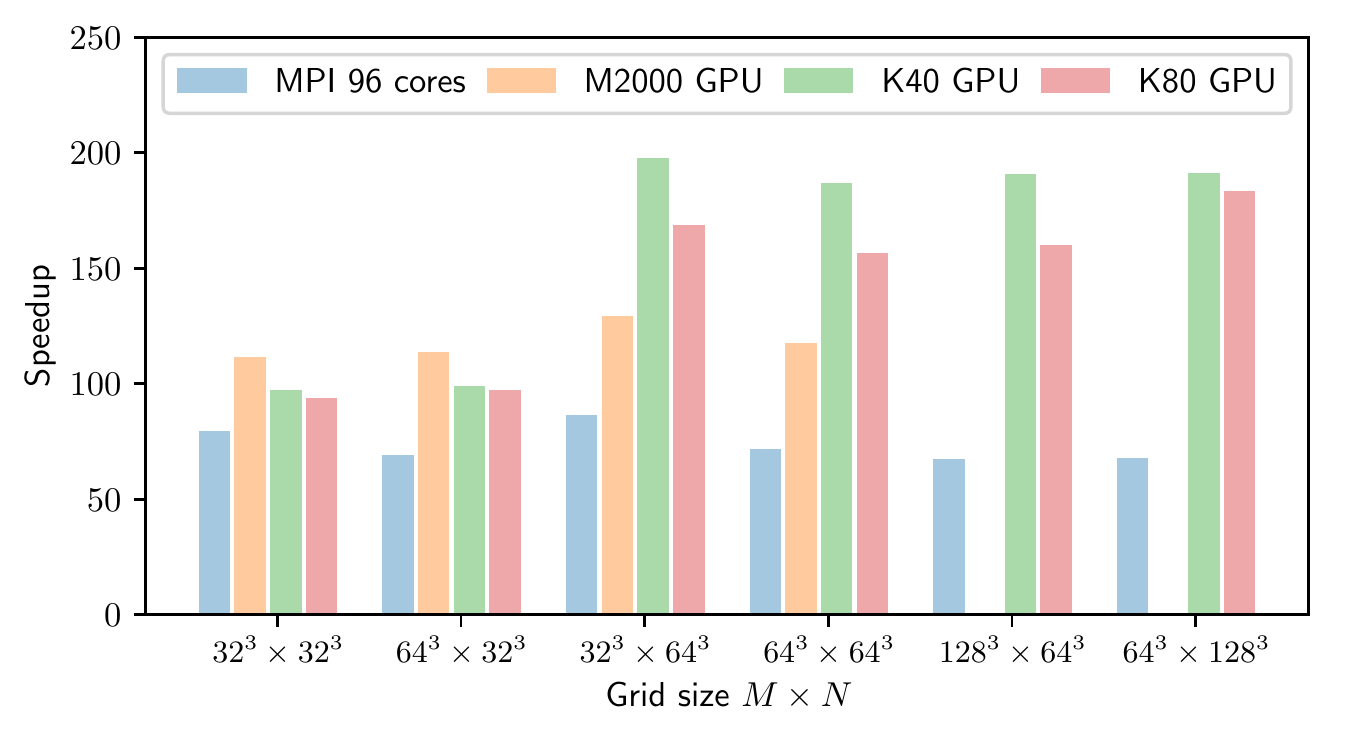}
	\caption{Speedup for the lid-driven cavity flow case with various grid sizes. The data of the M2000 GPU on the two largest grid sizes are not available due to insufficient global memory.}\label{fig:speed_up}
\end{figure}

\subsection{Kernel performance analysis}
We now analyze the performance of the two major kernels, i.e. \texttt{sweepSlice} and  \texttt{momentSlice} (Algorithms~\ref{alg:sweep_kernel} and~\ref{alg:moment_kernel}), in the GPU program using the lid-driven cavity flow as an example. The Nvidia visual profiler (nvvp)~\cite{nvidiacorporationProfilerUserGuide2018} is employed to measure various runtime metrics of the kernels on different GPUs when simulating the lid-driven cavity flow. The general metrics are shown in Table~\ref{tab:metrics} for the case with the grid sizes of $64^3\times 64^3$, $128^3\times 64^3$ and $64^3\times 128^3 $. These metrics indicate how efficient the various GPU resources are being used by the kernel functions, and reveal the performance limiters, e.g. memory bandwidth band, compute band or instruction/memory latencies. The relative runtime in the table is the ratio the kernel execution time to the overall computing time for a single iteration step.

The results in Table~\ref{tab:metrics} show that for the $64^3$ velocity grid size, the \texttt{sweepSlice} kernel takes slight longer time than the \texttt{momentSlice} kernel. While the situation reverses for the velocity grid size of $128^3$. This means the  \texttt{momentSlice} kernel does not scale equally well as the \texttt{sweepSlice} kernel.

The \texttt{sweepSlice} kernel's SM occupancy on M2000 and K40 is around $61\%$ while it is $96\%$ on the K80 GPU. The relative lower occupancy on M2000 and K40 GPU is due to the fewer registers available on their SM (64K as opposed to 128K on K80, see Table~\ref{tab:gpu_specs}). The profiling shows that the GPU performance on M2000 and K40 is limited by the instruction and memory fetch latencies, but when we restrict the kernel to use fewer registers via compiler flag, the achieved occupancy can be increased to $69\%$ but the overall computing time actually increases due to severer instruction and memory dependencies. The \texttt{sweepSlice} kernel's performance on K80 is bounded by the computing, i.e., the instruction execution speed. A further investigation using the profiler reveals that the load/store and arithmetic function units utilizations on the K80 GPU already reach a medium to high level.

In the \texttt{momentSlice} kernel, each thread only requires 32 registers but each thread block needs 8208 bytes of share memory.  On the K40 GPU, The SM occupancy is limited by the share memory size which only has 48 KB per SM. While on the M2000 and K80 GPUs, which have more share memory (see Table~\ref{tab:gpu_specs}), the SMs are almost fully occupied. On K40 and K80, the performance is limited by shared memory bandwidths measured at 1.5 TB/s and 1.3 TB/s respectively, which are close to the devices' limits.

We also investigate the effect of workload on the kernels' performance on K40 and K80 with larger grid sizes: $64^3\times 128^3$ and $128^3 \times  64^3$. Although a velocity grid size as large as $128^3$ is too expensive for most practical applications and should be avoided with more sophisticated velocity grids such as non-uniform and adaptive velocity grids, and conservative moment-evaluation procedures, we use the large velocity grid as an extreme case to explore the performance limit of the kernel functions under the condition of sufficiently abundant parallelism. For example, with a velocity grid size of $128^3$, each SM on the K40 GPU can be populated with at least 68 thread blocks which would be enough to keep the SMs busy and minimize the trail effect (some SMs get one thread block less than others due to the round-robin scheduling policy and the number of SM not dividing the total number of thread blocks).

{\renewcommand{\arraystretch}{1.2}%
	\begin{table}[htbp]
		\centering
		\caption{Utilization metrics of GPU resource for the \texttt{sweepSlice} and \texttt{momentSlice} kernels. The data are retrieved from Nvidia Visual Profiler (nvvp). The test case is the lid-driven cavity flow with the grid size of $64^3\times 64^3$.  The \emph{latency} in the table stands for instruction and memory latencies.}
		\begin{tabular}{lrrrrrrr}
			\toprule
			Kernel                                              &   GPU &  Achieved &     Compute &      Memory & Device memory & Performance & Relative \\
			& model & occupancy & utilization & utilization &    throughput &     limiter &  runtime \\ \midrule
			\multicolumn{8}{c}{grid size $64^3\times 64^3$}                                                \\ \cline{2-8}
			\multirow{3}{*}{\texttt{sweepSlice}}               & M2000 &  $60.1\%$ &      $64\%$ &      $55\%$ &   $53.4$ GB/s &     latency & $58.0\%$ \\
			&   K40 &  $61.6\%$ &      $65\%$ &      $38\%$ &  $112.5$ GB/s &     latency & $51.6\%$ \\
			&   K80 &  $95.5\%$ &      $78\%$ &      $46\%$ &   $90.4$ GB/s &     compute & $53.9\%$ \\ \hline
			\vspace{2pt}
			
			\multirow{3}{*}{\texttt{momentSlice}} & M2000 &  $99.7\%$ &      $42\%$ &      $56\%$ &   $52.5$ GB/s &     latency & $42.0\%$ \\
			&   K40 &  $61.3\%$ &      $30\%$ &      $78\%$ &   $84.2$ GB/s &      memory & $48.4\%$ \\
			&   K80 &  $98.4\%$ &      $31\%$ &      $76\%$ &   $73.7$ GB/s &      memory & $46.1\%$ \\ \hline
			\multicolumn{8}{c}{grid size $128^3\times 64^3$}                                               \\ \cline{2-8}
			\multirow{2}{*}{\texttt{sweepSlice}}               &   K40 &  $61.6\%$ &      $59\%$ &      $47\%$ &  $128.8$ GB/s &     latency & $51.4\%$ \\
			&   K80 &  $95.6\%$ &      $79\%$ &      $46\%$ &   $91.6$ GB/s &     compute & $53.6\%$ \\ \hline
			\vspace{2pt}
			
			\multirow{2}{*}{\texttt{momentSlice}} &   K40 &  $61.4\%$ &      $31\%$ &      $75\%$ &   $84.0$ GB/s &      memory & $48.6\%$ \\
			&   K80 &  $98.4\%$ &      $30\%$ &      $76\%$ &   $73.8$ GB/s &      memory & $46.4\%$ \\
			\multicolumn{8}{c}{grid size $64^3\times 128^3$}                                               \\ \cline{2-8}
			\multirow{2}{*}{\texttt{sweepSlice}}               &   K40 &  $62.4\%$ &      $58\%$ &      $48\%$ &  $143.1$ GB/s &     latency & $48.0\%$ \\
			&   K80 &  $98.7\%$ &      $78\%$ &      $58\%$ &  $139.4$ GB/s &      compute & $43.1\%$ \\ \hline
			\multirow{2}{*}{\texttt{momentSlice}}                  &   K40 &  $62.1\%$ &      $29\%$ &      $78\%$ &  $163.3$ GB/s &      memory & $52.0\%$ \\
			&   K80 &  $99.4\%$ &      $30\%$ &      $77\%$ &  $157.8$ GB/s &      memory & $56.9\%$ \\ \bottomrule
			
		\end{tabular}%
		\label{tab:metrics}%
	\end{table}%
	
	\section{Conclusion and discussions}\label{sec:conlusions}
	In summary, an efficient memory-reduced GPU accelerated iterative DVM for kinetic model equations has been developed.  Different from previously reported GPU accelerations of explicit kinetic schemes, the current implementation is based on a fast converging iterative scheme. The memory reduction techniques in both the molecular velocity and physical spaces reduce memory requirements significantly from terabytes to gigabytes. The test cases including a supersonic rarefied flow past a cube with the grid points of the phase space up to 0.7 trillion demonstrated the capability of the proposed GPU algorithm in simulating large-scale 3D flows on a single GPU. The performance profiling on different GPUs show the implementations of our GPU kernel functions have good utilization levels of GPU resources which can achieve nearly a two-hundred speedup on a Tesla K40 against its serial CPU counterpart.
	
	The performance of the current implementation benefits from a large velocity grid as $64^3$ or more, where there is enough parallelism to realize the full potential of high-end Tesla series GPUs with thousands of cores.  However, the current trend in the latest released GPUs is that the number of cores is increasing even more, e.g., 5120 on the V100 GPU released in 2017. In such case, to fully utilize the improved computing power in the future, more parallelism in the algorithm should be applied other than using only the molecular-velocity-space parallelization. Another possible improvement in the future is to extend the first order upwind scheme to second order by storing three consecutive slices of cell's distribution functions.
	
	\section*{Acknowledgments}
	This project leading to this paper has received funding from the European Union’s Horizon 2020 research and innovation programme under the Marie Skłodowska-Curie grant agreement number 793007. Financial support from the UK Engineering and Physical Sciences Research Council (EPSRC) under grant no. EP/M021475/1 is gratefully acknowledged. S.~Chen acknowledge the financial support from National Science Foundation of China (Grants No. 91530319). Computing time during the program development \& testing on the ARCHER is provided by the “UK Consortium on Mesoscale Engineering Sciences (UKCOMES)” under the UK EPSRC grant no. EP/L00030X/1.
	
	\biboptions{square,comma,compress}
	\bibliographystyle{elsarticle-num}
	\bibliography{implicit_gpu}

\begin{thebibliography}{10}
\expandafter\ifx\csname url\endcsname\relax
  \def\url#1{\texttt{#1}}\fi
\expandafter\ifx\csname urlprefix\endcsname\relax\def\urlprefix{URL }\fi
\expandafter\ifx\csname href\endcsname\relax
  \def\href#1#2{#2} \def\path#1{#1}\fi

\bibitem{birdMolecularGasDynamics1994}
G.~A. Bird, Molecular {{Gas Dynamics}} and the {{Direct Simulation}} of {{Gas
  Flows}}, {Clarendon Press}, 1994.

\bibitem{birdDSMCMethod2013}
G.~A. Bird, The {{DSMC Method}}, {CreateSpace Independent Publishing Platform},
  2013.

\bibitem{broadwellStudyRarefiedShear1964}
J.~E. Broadwell, Study of rarefied shear flow by the discrete velocity method,
  J. Fluid Mech. 19 (1964) 401--414.
\newblock \href {http://dx.doi.org/10.1017/S0022112064000817}
  {\path{doi:10.1017/S0022112064000817}}.

\bibitem{yangRarefiedFlowComputations1995}
J.~Y. Yang, J.~C. Huang, Rarefied flow computations using nonlinear model
  {{Boltzmann}} equations, J. Comput. Phys. 120 (1995) 323--339.
\newblock \href {http://dx.doi.org/10.1006/jcph.1995.1168}
  {\path{doi:10.1006/jcph.1995.1168}}.

\bibitem{aristovDirectMethodsSolving2001}
V.~V. Aristov, Direct Methods for Solving the {{Boltzmann}} Equation and Study
  of Nonequilibrium Flows, {Springer Science \& Business Media}, 2001.

\bibitem{mieussensSurveyDeterministicSolvers2014a}
L.~Mieussens, A survey of deterministic solvers for rarefied flows
  ({{Invited}}), in: {{AIP Conference Proceedings}}, Vol. 1628, 2014, pp.
  943--951.
\newblock \href {http://dx.doi.org/10.1063/1.4902695}
  {\path{doi:10.1063/1.4902695}}.

\bibitem{soneMolecularGasDynamics2007}
Y.~Sone, Molecular {{Gas Dynamics}}: {{Theory}}, {{Techniques}}, and
  {{Applications}}, {Springer Science \& Business Media.}, 2007.

\bibitem{wuSolvingBoltzmannEquation2014}
L.~Wu, J.~M. Reese, Y.~Zhang, Solving the {{Boltzmann}} equation
  deterministically by the fast spectral method: Application to gas microflows,
  J. Fluid Mech. 746 (2014) 53--84.
\newblock \href {http://dx.doi.org/10.1017/jfm.2014.79}
  {\path{doi:10.1017/jfm.2014.79}}.

\bibitem{mengLatticeEllipsoidalStatistical2013}
J.~Meng, Y.~Zhang, N.~G. Hadjiconstantinou, G.~A. Radtke, X.~Shan, Lattice
  ellipsoidal statistical {{BGK}} model for thermal non-equilibrium flows, J.
  Fluid Mech. 718 (2013) 347--370.
\newblock \href {http://dx.doi.org/10.1017/jfm.2012.616}
  {\path{doi:10.1017/jfm.2012.616}}.

\bibitem{huangUnifiedGasKineticScheme2013}
J.-C. Huang, K.~Xu, P.~Yu, A {{Unified Gas}}-{{Kinetic Scheme}} for
  {{Continuum}} and {{Rarefied Flows III}}: {{Microflow Simulations}}, Commun.
  Comput. Phys. 14 (2013) 1147--1173.
\newblock \href {http://dx.doi.org/10.4208/cicp.190912.080213a}
  {\path{doi:10.4208/cicp.190912.080213a}}.

\bibitem{wuApparentPermeabilityPorous2017}
L.~Wu, M.~T. Ho, L.~Germanou, X.-J. Gu, C.~Liu, K.~Xu, Y.~Zhang, On the
  apparent permeability of porous media in rarefied gas flows, J. Fluid Mech.
  822 (2017) 398--417.
\newblock \href {http://dx.doi.org/10.1017/jfm.2017.300}
  {\path{doi:10.1017/jfm.2017.300}}.

\bibitem{jinAsymptoticPreservingAP2010}
S.~Jin, Asymptotic preserving ({{AP}}) schemes for multiscale kinetic and
  hyperbolic equations: A review, Lect. Notes Summer Sch. ``Methods Models
  Kinet. Theory''MMKT Porto Ercole Grosseto Italy (2010) 177--216.

\bibitem{xuUnifiedGaskineticScheme2010}
K.~Xu, J.-C. Huang, A unified gas-kinetic scheme for continuum and rarefied
  flows, J. Comput. Phys. 229 (2010) 7747--7764.
\newblock \href {http://dx.doi.org/10.1016/j.jcp.2010.06.032}
  {\path{doi:10.1016/j.jcp.2010.06.032}}.

\bibitem{mieussensAsymptoticPreservingProperty2013}
L.~Mieussens, On the asymptotic preserving property of the unified gas kinetic
  scheme for the diffusion limit of linear kinetic models, J. Comput. Phys. 253
  (2013) 138--156.
\newblock \href {http://dx.doi.org/10.1016/j.jcp.2013.07.002}
  {\path{doi:10.1016/j.jcp.2013.07.002}}.

\bibitem{guoDiscreteUnifiedGas2013}
Z.~Guo, K.~Xu, R.~Wang, Discrete unified gas kinetic scheme for all {{Knudsen}}
  number flows: {{Low}}-speed isothermal case, Phys. Rev. E 88 (2013) 033305.
\newblock \href {http://dx.doi.org/10.1103/PhysRevE.88.033305}
  {\path{doi:10.1103/PhysRevE.88.033305}}.

\bibitem{guoDiscreteUnifiedGas2015}
Z.~Guo, R.~Wang, K.~Xu, Discrete unified gas kinetic scheme for all {{Knudsen}}
  number flows. {{II}}. {{Thermal}} compressible case, Phys. Rev. E 91 (2015)
  033313.
\newblock \href {http://dx.doi.org/10.1103/PhysRevE.91.033313}
  {\path{doi:10.1103/PhysRevE.91.033313}}.

\bibitem{liuUnifiedGaskineticScheme2014a}
S.~Liu, P.~Yu, K.~Xu, C.~Zhong, Unified gas-kinetic scheme for diatomic
  molecular simulations in all flow regimes, J. Comput. Phys. 259 (2014)
  96--113.
\newblock \href {http://dx.doi.org/10.1016/j.jcp.2013.11.030}
  {\path{doi:10.1016/j.jcp.2013.11.030}}.

\bibitem{xuDirectModelingComputational2015}
K.~Xu, Direct {{Modeling}} for {{Computational Fluid Dynamics}}:
  {{Construction}} and {{Application}} of {{Unified Gas}}-{{Kinetic Schemes}},
  Advances in Computational Fluid Dynamics, {World Scientific Publishing},
  2015.

\bibitem{zhuDiscreteUnifiedGas2016}
L.~Zhu, Z.~Guo, K.~Xu, Discrete unified gas kinetic scheme on unstructured
  meshes, Comput. Fluids 127 (2016) 211--225.
\newblock \href {http://dx.doi.org/10.1016/j.compfluid.2016.01.006}
  {\path{doi:10.1016/j.compfluid.2016.01.006}}.

\bibitem{zhuApplicationDiscreteUnified2017}
L.~Zhu, Z.~Guo, Application of discrete unified gas kinetic scheme to thermally
  induced nonequilibrium flows, Comput. Fluids\href
  {http://dx.doi.org/10.1016/j.compfluid.2017.09.019}
  {\path{doi:10.1016/j.compfluid.2017.09.019}}.

\bibitem{mieussensDiscreteVelocityModel2000}
L.~Mieussens, Discrete velocity model and implicit scheme for the bgk equation
  of rarefied gas dynamics, Math. Models Methods Appl. Sci. 10 (2000)
  1121--1149.
\newblock \href {http://dx.doi.org/10.1142/S0218202500000562}
  {\path{doi:10.1142/S0218202500000562}}.

\bibitem{titarevConstructionComparisonParallel2014}
V.~Titarev, M.~Dumbser, S.~Utyuzhnikov, Construction and comparison of parallel
  implicit kinetic solvers in three spatial dimensions, J. Comput. Phys. 256
  (2014) 17--33.
\newblock \href {http://dx.doi.org/10.1016/j.jcp.2013.08.051}
  {\path{doi:10.1016/j.jcp.2013.08.051}}.

\bibitem{zhuImplicitUnifiedGaskinetic2016}
Y.~Zhu, C.~Zhong, K.~Xu, Implicit unified gas-kinetic scheme for steady state
  solutions in all flow regimes, J. Comput. Phys. 315 (2016) 16--38.
\newblock \href {http://dx.doi.org/10.1016/j.jcp.2016.03.038}
  {\path{doi:10.1016/j.jcp.2016.03.038}}.

\bibitem{zhuUnifiedGaskineticScheme2017}
Y.~Zhu, C.~Zhong, K.~Xu, Unified gas-kinetic scheme with multigrid convergence
  for rarefied flow study, Phys. Fluids 29 (2017) 096102.
\newblock \href {http://dx.doi.org/10.1063/1.4994020}
  {\path{doi:10.1063/1.4994020}}.

\bibitem{yangImplicitSchemeMemory2018}
L.~M. Yang, C.~Shu, W.~M. Yang, J.~Wu, An implicit scheme with memory reduction
  technique for steady state solutions of {{DVBE}} in all flow regimes, Phys.
  Fluids 30 (2018) 040901.
\newblock \href {http://dx.doi.org/10.1063/1.5008479}
  {\path{doi:10.1063/1.5008479}}.

\bibitem{wangComparativeStudyDiscrete2018a}
P.~Wang, M.~T. Ho, L.~Wu, Z.~Guo, Y.~Zhang, A comparative study of discrete
  velocity methods for low-speed rarefied gas flows, Computers \& Fluids 161
  (2018) 33--46.
\newblock \href {http://dx.doi.org/10.1016/j.compfluid.2017.11.006}
  {\path{doi:10.1016/j.compfluid.2017.11.006}}.

\bibitem{filbetHighOrderNumerical2003}
F.~Filbet, G.~Russo, High order numerical methods for the space non-homogeneous
  {{Boltzmann}} equation, J. Comput. Phys. 186 (2003) 457--480.
\newblock \href {http://dx.doi.org/10.1016/S0021-9991(03)00065-2}
  {\path{doi:10.1016/S0021-9991(03)00065-2}}.

\bibitem{alexeenkoHighOrderDiscontinuousGalerkin2008}
A.~Alexeenko, C.~Galitzine, A.~Alekseenko, High-{{Order Discontinuous Galerkin
  Method}} for {{Boltzmann Model Equations}}, in: 40th {{Thermophysics
  Conference}}, {American Institute of Aeronautics and Astronautics}, Seattle,
  Washington, 2008, p. 4256.
\newblock \href {http://dx.doi.org/10.2514/6.2008-4256}
  {\path{doi:10.2514/6.2008-4256}}.

\bibitem{wuThirdorderDiscreteUnified2018}
C.~Wu, B.~Shi, C.~Shu, Z.~Chen, Third-order discrete unified gas kinetic scheme
  for continuum and rarefied flows: {{Low}}-speed isothermal case, Phys. Rev. E
  97 (2018) 023306.
\newblock \href {http://dx.doi.org/10.1103/PhysRevE.97.023306}
  {\path{doi:10.1103/PhysRevE.97.023306}}.

\bibitem{suHighorderHybridizableDiscontinuous2018}
W.~Su, P.~Wang, Y.~Zhang, L.~Wu, A high-order hybridizable discontinuous
  {{Galerkin}} method with fast convergence to steady-state solutions of the
  gas kinetic equation, J. Comput. Phys.\href
  {http://dx.doi.org/10.1016/j.jcp.2018.08.050}
  {\path{doi:10.1016/j.jcp.2018.08.050}}.

\bibitem{liGaskineticNumericalStudies2009}
Z.-H. Li, H.-X. Zhang, Gas-kinetic numerical studies of three-dimensional
  complex flows on spacecraft re-entry, J. Comput. Phys. 228 (2009) 1116--1138.
\newblock \href {http://dx.doi.org/10.1016/j.jcp.2008.10.013}
  {\path{doi:10.1016/j.jcp.2008.10.013}}.

\bibitem{liRarefiedGasFlow2015}
Z.-H. Li, A.-P. Peng, H.-X. Zhang, J.-Y. Yang, Rarefied gas flow simulations
  using high-order gas-kinetic unified algorithms for {{Boltzmann}} model
  equations, Prog. Aerosp. Sci. 74 (2015) 81--113.
\newblock \href {http://dx.doi.org/10.1016/j.paerosci.2014.12.002}
  {\path{doi:10.1016/j.paerosci.2014.12.002}}.

\bibitem{dimarcoUltraEfficientKinetic2015}
G.~Dimarco, R.~Loub\`ere, J.~Narski, Towards an ultra efficient kinetic scheme.
  {{Part III}}: {{High}}-performance-computing, J. Comput. Phys. 284 (2015)
  22--39.
\newblock \href {http://dx.doi.org/10.1016/j.jcp.2014.12.023}
  {\path{doi:10.1016/j.jcp.2014.12.023}}.

\bibitem{liHighPerformanceParallel2016}
S.~Li, Q.~Li, S.~Fu, J.~Xu, The high performance parallel algorithm for
  {{Unified Gas}}-{{Kinetic Scheme}}, in: {{AIP Conference Proceedings}},
  Victoria, BC, Canada, 2016, p. 180007.
\newblock \href {http://dx.doi.org/10.1063/1.4967676}
  {\path{doi:10.1063/1.4967676}}.

\bibitem{titarevApplicationModelKinetic2017}
V.~A. Titarev, Application of model kinetic equations to hypersonic rarefied
  gas flows, Comput. Fluids\href
  {http://dx.doi.org/10.1016/j.compfluid.2017.06.019}
  {\path{doi:10.1016/j.compfluid.2017.06.019}}.

\bibitem{dimarcoEfficientNumericalMethod2018}
G.~Dimarco, R.~Loub\`ere, J.~Narski, T.~Rey, An efficient numerical method for
  solving the {{Boltzmann}} equation in multidimensions, J. Comput. Phys. 353
  (2018) 46--81.
\newblock \href {http://dx.doi.org/10.1016/j.jcp.2017.10.010}
  {\path{doi:10.1016/j.jcp.2017.10.010}}.

\bibitem{hoMultilevelParallelSolver2018}
M.~T. Ho, L.~Zhu, L.~Wu, P.~Wang, Z.~Guo, Z.-H. Li, Y.~Zhang, A multi-level
  parallel solver for rarefied gas flows in porous media, Comput. Phys.
  Commun.\href {http://dx.doi.org/10.1016/j.cpc.2018.08.009}
  {\path{doi:10.1016/j.cpc.2018.08.009}}.

\bibitem{kolobovUnifiedSolverRarefied2007}
V.~I. Kolobov, R.~R. Arslanbekov, V.~V. Aristov, A.~A. Frolova, S.~A. Zabelok,
  Unified solver for rarefied and continuum flows with adaptive mesh and
  algorithm refinement, J. Comput. Phys. 223 (2007) 589--608.
\newblock \href {http://dx.doi.org/10.1016/j.jcp.2006.09.021}
  {\path{doi:10.1016/j.jcp.2006.09.021}}.

\bibitem{chenUnifiedGasKinetic2012}
S.~Chen, K.~Xu, C.~Lee, Q.~Cai, A unified gas kinetic scheme with moving mesh
  and velocity space adaptation, J. Comput. Phys. 231 (2012) 6643--6664.
\newblock \href {http://dx.doi.org/10.1016/j.jcp.2012.05.019}
  {\path{doi:10.1016/j.jcp.2012.05.019}}.

\bibitem{barangerLocallyRefinedDiscrete2014}
C.~Baranger, J.~Claudel, N.~H\'erouard, L.~Mieussens, Locally refined discrete
  velocity grids for stationary rarefied flow simulations, J. Comput. Phys.
  257, Part A (2014) 572--593.
\newblock \href {http://dx.doi.org/10.1016/j.jcp.2013.10.014}
  {\path{doi:10.1016/j.jcp.2013.10.014}}.

\bibitem{brullLocalDiscreteVelocity2014}
S.~Brull, L.~Mieussens, Local discrete velocity grids for deterministic
  rarefied flow simulations, J. Comput. Phys. 266 (2014) 22--46.
\newblock \href {http://dx.doi.org/10.1016/j.jcp.2014.01.050}
  {\path{doi:10.1016/j.jcp.2014.01.050}}.

\bibitem{zabelokAdaptiveKineticfluidSolvers2015}
S.~Zabelok, R.~Arslanbekov, V.~Kolobov, Adaptive kinetic-fluid solvers for
  heterogeneous computing architectures, J. Comput. Phys. 303 (2015) 455--469.
\newblock \href {http://dx.doi.org/10.1016/j.jcp.2015.10.003}
  {\path{doi:10.1016/j.jcp.2015.10.003}}.

\bibitem{fanGPUClusterHigh2004}
Z.~Fan, F.~Qiu, A.~Kaufman, S.~{Yoakum-Stover}, {{GPU Cluster}} for {{High
  Performance Computing}}, in: Proceedings of the 2004 {{ACM}}/{{IEEE
  Conference}} on {{Supercomputing}}, SC '04, {IEEE Computer Society},
  Washington, DC, USA, 2004, pp. 47--.
\newblock \href {http://dx.doi.org/10.1109/SC.2004.26}
  {\path{doi:10.1109/SC.2004.26}}.

\bibitem{tolkeTeraFLOPComputingDesktop2008}
J.~T\"olke, M.~Krafczyk, {{TeraFLOP}} computing on a desktop {{PC}} with
  {{GPUs}} for {{3D CFD}}, Int. J. Numer. Methods Fluids 22 (2008) 443--456.
\newblock \href {http://dx.doi.org/10.1080/10618560802238275}
  {\path{doi:10.1080/10618560802238275}}.

\bibitem{obrechtNewApproachLattice2011}
C.~Obrecht, F.~Kuznik, B.~Tourancheau, J.-J. Roux, A new approach to the
  lattice {{Boltzmann}} method for graphics processing units, Computers \&
  Mathematics with Applications 61 (2011) 3628--3638.
\newblock \href {http://dx.doi.org/10.1016/j.camwa.2010.01.054}
  {\path{doi:10.1016/j.camwa.2010.01.054}}.

\bibitem{mcclureNovelHeterogeneousAlgorithm2014}
J.~E. McClure, J.~F. Prins, C.~T. Miller, A novel heterogeneous algorithm to
  simulate multiphase flow in porous media on multicore
  {{CPU}}\textendash{{GPU}} systems, Comput. Phys. Commun. 185 (2014)
  1865--1874.
\newblock \href {http://dx.doi.org/10.1016/j.cpc.2014.03.012}
  {\path{doi:10.1016/j.cpc.2014.03.012}}.

\bibitem{xuAcceleratedLatticeBoltzmann2017}
A.~Xu, L.~Shi, T.~S. Zhao, Accelerated lattice {{Boltzmann}} simulation using
  {{GPU}} and {{OpenACC}} with data management, Int. J. Heat Mass Transfer 109
  (2017) 577--588.
\newblock \href {http://dx.doi.org/10.1016/j.ijheatmasstransfer.2017.02.032}
  {\path{doi:10.1016/j.ijheatmasstransfer.2017.02.032}}.

\bibitem{suLargescaleSimulationsMultiple2012}
C.~C. Su, M.~R. Smith, F.~A. Kuo, J.~S. Wu, C.~W. Hsieh, K.~C. Tseng,
  Large-scale simulations on multiple {{Graphics Processing Units}} ({{GPUs}})
  for the direct simulation {{Monte Carlo}} method, J. Comput. Phys. 231 (2012)
  7932--7958.
\newblock \href {http://dx.doi.org/10.1016/j.jcp.2012.07.038}
  {\path{doi:10.1016/j.jcp.2012.07.038}}.

\bibitem{goldsworthyGPUCUDABased2014}
M.~J. Goldsworthy, A {{GPU}}\textendash{{CUDA}} based direct simulation {{Monte
  Carlo}} algorithm for real gas flows, Comput. Fluids 94 (2014) 58--68.
\newblock \href {http://dx.doi.org/10.1016/j.compfluid.2014.01.033}
  {\path{doi:10.1016/j.compfluid.2014.01.033}}.

\bibitem{jambunathanCHAOSOctreebasedPICDSMC2018}
R.~Jambunathan, D.~A. Levin, {{CHAOS}}: {{An}} octree-based {{PIC}}-{{DSMC}}
  code for modeling of electron kinetic properties in a plasma plume using
  {{MPI}}-{{CUDA}} parallelization, J. Comput. Phys. 373 (2018) 571--604.
\newblock \href {http://dx.doi.org/10.1016/j.jcp.2018.07.005}
  {\path{doi:10.1016/j.jcp.2018.07.005}}.

\bibitem{frezzottiSolvingModelKinetic2011}
A.~Frezzotti, G.~P. Ghiroldi, L.~Gibelli, Solving model kinetic equations on
  {{GPUs}}, Comput. Fluids 50 (2011) 136--146.
\newblock \href {http://dx.doi.org/10.1016/j.compfluid.2011.07.004}
  {\path{doi:10.1016/j.compfluid.2011.07.004}}.

\bibitem{frezzottiSolvingBoltzmannEquation2011}
A.~Frezzotti, G.~P. Ghiroldi, L.~Gibelli, Solving the {{Boltzmann}} equation on
  {{GPUs}}, Comput. Phys. Commun. 182 (2011) 2445--2453.
\newblock \href {http://dx.doi.org/10.1016/j.cpc.2011.07.002}
  {\path{doi:10.1016/j.cpc.2011.07.002}}.

\bibitem{frezzottiDirectSolutionBoltzmann2011}
A.~Frezzotti, G.~P. Ghiroldi, L.~Gibelli, Direct solution of the {{Boltzmann}}
  equation for a binary mixture on {{GPUs}}, in: {{AIP Conference
  Proceedings}}, Vol. 1333, 2011, pp. 884--889.
\newblock \href {http://dx.doi.org/10.1063/1.3562757}
  {\path{doi:10.1063/1.3562757}}.

\bibitem{klossSolvingBoltzmannEquation2010}
Y.~Y. Kloss, P.~V. Shuvalov, F.~G. Tcheremissine, Solving {{Boltzmann}}
  equation on {{GPU}}, Procedia Comput. Sci. 1 (2010) 1083--1091.
\newblock \href {http://dx.doi.org/j.procs.2010.04.120}
  {\path{doi:j.procs.2010.04.120}}.

\bibitem{aristovAccelerationDeterministicBoltzmann2011}
V.~V. Aristov, A.~A. Frolova, S.~A. Zabelok, V.~I. Kolobov, R.~R. Arslanbekov,
  Acceleration of {{Deterministic Boltzmann Solver}} with {{Graphics Processing
  Units}}, in: {{AIP Conference Proceedings}}, Vol. 1333, Pacific Grove,
  California, (USA), 2011, pp. 867--872.
\newblock \href {http://dx.doi.org/10.1063/1.3562754}
  {\path{doi:10.1063/1.3562754}}.

\bibitem{zabelokGPUAcceleratedKinetic2012}
S.~A. Zabelok, V.~I. Kolobov, R.~R. Arslanbekov, {{GPU Accelerated Kinetic
  Solvers}} for {{Rarefied Gas Dynamics}}, in: {{AIP Conference Proceedings}},
  Vol. 1501, 2012, pp. 429--434.
\newblock \href {http://dx.doi.org/10.1063/1.4769562}
  {\path{doi:10.1063/1.4769562}}.

\bibitem{rovenskayaNumericalInvestigationEffect2015}
O.~Rovenskaya, G.~Croce, Numerical investigation of the effect of boundary
  conditions for a highly rarefied gas flow using the {{GPU}} accelerated
  {{Boltzmann}} solver, Comput. Fluids 110 (2015) 77--87.
\newblock \href {http://dx.doi.org/10.1016/j.compfluid.2014.10.015}
  {\path{doi:10.1016/j.compfluid.2014.10.015}}.

\bibitem{zabelokMultiGPUKineticSolvers2014}
S.~Zabelok, R.~Arslanbekov, V.~Kolobov, Multi-{{GPU Kinetic Solvers}} using
  {{MPI}} and {{CUDA}}, in: {{AIP Conference Proceedings}}, Vol. 1628, 2014,
  pp. 539--546.
\newblock \href {http://dx.doi.org/10.1063/1.4902640}
  {\path{doi:10.1063/1.4902640}}.

\bibitem{shakhovGeneralizationKrookKinetic1968}
E.~M. Shakhov, Generalization of the {{Krook}} kinetic relaxation equation,
  Fluid Dyn. 3 (1968) 95--96.
\newblock \href {http://dx.doi.org/10.1007/BF01029546}
  {\path{doi:10.1007/BF01029546}}.

\bibitem{chenUnifiedImplicitScheme2017}
S.~Chen, C.~Zhang, L.~Zhu, Z.~Guo, A unified implicit scheme for kinetic model
  equations. {{Part I}}. {{Memory}} reduction technique, Sci. Bull. 62 (2017)
  119--129.
\newblock \href {http://dx.doi.org/10.1016/j.scib.2016.12.010}
  {\path{doi:10.1016/j.scib.2016.12.010}}.

\bibitem{zhuThermallyInducedRarefied2017}
L.~Zhu, X.~Yang, Z.~Guo, Thermally induced rarefied gas flow in a
  three-dimensional enclosure with square cross-section, Phys. Rev. Fluids 2
  (2017) 123402.
\newblock \href {http://dx.doi.org/10.1103/PhysRevFluids.2.123402}
  {\path{doi:10.1103/PhysRevFluids.2.123402}}.

\bibitem{sharipovRarefiedGasDynamics2015}
F.~Sharipov, Rarefied Gas Dynamics: Fundamentals for Research and Practice,
  {John Wiley \& Sons}, 2015.

\bibitem{wassermanPerformanceScalabilityAnalysis2000}
H.~Wasserman, A.~Hoisie, A.~Hoisie, O.~Lubeck, O.~Lubeck, Performance and
  {{Scalability Analysis}} of {{Teraflop}}-{{Scale Parallel Architectures}}
  using {{Multidimensional Wavefront Applications}}, Int. J. High Perform.
  Comput. Appl. 14 (2000) 330--346.
\newblock \href {http://dx.doi.org/0.1177/109434200001400405}
  {\path{doi:0.1177/109434200001400405}}.

\bibitem{moustafaSharedMemoryParallelism2015}
S.~Moustafa, I.~{Dutka-Malen}, L.~Plagne, A.~Pon{\c c}ot, P.~Ramet, Shared
  memory parallelism for {{3D Cartesian}} discrete ordinates solver, Ann. Nucl.
  Energy 82 (2015) 179--187.
\newblock \href {http://dx.doi.org/10.1016/j.anucene.2014.08.034}
  {\path{doi:10.1016/j.anucene.2014.08.034}}.

\bibitem{deakinImprovedParallelismScheme2016}
T.~Deakin, S.~{McIntosh-Smith}, M.~Martineau, W.~Gaudin, An improved
  parallelism scheme for deterministic discrete ordinates transport, Int. J.
  High Perform. Comput. Appl. 32 (2016) 555--569.
\newblock \href {http://dx.doi.org/10.1177/1094342016668978}
  {\path{doi:10.1177/1094342016668978}}.

\bibitem{scanlonOpenSourceParallel2010}
T.~J. Scanlon, E.~Roohi, C.~White, M.~Darbandi, J.~M. Reese, An open source,
  parallel {{DSMC}} code for rarefied gas flows in arbitrary geometries,
  Comput. Fluids 39 (2010) 2078--2089.
\newblock \href {http://dx.doi.org/10.1016/j.compfluid.2010.07.014}
  {\path{doi:10.1016/j.compfluid.2010.07.014}}.

\bibitem{nvidiacorporationProfilerUserGuide2018}
N.~Corporation,
  \href{https://docs.nvidia.com/cuda/profiler-users-guide/index.html}{Profiler
  {{User}}'s {{Guide}} v9.2} (2018).
\newline\urlprefix\url{https://docs.nvidia.com/cuda/profiler-users-guide/index.html}

\end{thebibliography}
\end{document}